\documentclass[twocolumn,superscriptaddress,amsmath, amssymb, amsfonts,preprintnumbers,aps,prd,longbibliography,nofootinbib]{revtex4-1}

\renewcommand{\sec}[1]{\textit{#1. --- }}

\usepackage{graphicx}
\usepackage{dcolumn}
\usepackage{bm}
 \usepackage{amstext}
 \usepackage{amssymb}
\usepackage{subcaption}
 \usepackage{amsmath}
 \usepackage{graphicx}
 \usepackage{color}
 \usepackage{bbold}
 \usepackage{delimset} 
\usepackage[colorlinks=true]{hyperref}

\usepackage{tikz}
\usetikzlibrary{decorations.pathmorphing}
\usetikzlibrary{decorations.markings}
\usetikzlibrary{positioning, shapes, snakes, arrows}

\tikzset{
	graviton/.style={line width=.8pt, -latex,decorate, decoration={snake, segment length=4pt,amplitude=1.8pt, pre length=.1cm, post length=.25cm}},
	worldline/.style={gray, line width=1pt},
	worldlineBold/.style={black, line width=.6pt},
	zUndirected/.style={line width=1pt},
	zParticle/.style={line width=1pt,postaction={decorate},decoration={markings,mark=at position .6 with {\arrow[#1]{latex}}}},
	zParticleF/.style={line width=1pt,postaction={decorate}},
	cscalar/.style={line width=1pt,postaction={decorate},decoration={markings,mark=at position .6 with {\arrow[#1]{latex}}}},
	cscalar2/.style={line width=1pt,postaction={decorate},decoration={markings,mark=at position .8 with {\arrow[#1]{latex}}}},
	photon/.style={line width =.8pt, decorate, decoration={snake, segment length=3pt, amplitude=1.8pt,  pre length=.1cm, post length=.1cm}}
}

\DeclareFontFamily{OT1}{pzc}{}
\DeclareFontShape{OT1}{pzc}{m}{it}{<-> s * [1.350] pzcmi7t}{}
\DeclareMathAlphabet{\mathpzc}{OT1}{pzc}{m}{it}

\def\cN{\mathcal{N}}
\def\cO{\mathcal{O}}

\def\cS{\mathcal{S}}

\def\eps{\epsilon}

\def\d{\mathrm{d}}
\def\D{\mathrm{D}}

\def\dd{\delta\!\!\!{}^-\!}

\def\d{\mathrm{d}}
\def\eps{\epsilon}

\def\braket#1{\langle #1 \rangle}

\def\nn{\nonumber}

\def\Eqn#1{Eq.~\eqref{#1}}

\def\Fig#1{Fig.~{\ref{#1}}}
\def\Figs#1#2{Figs.~{\ref{#1}} and~{\ref{#2}}}

\def\App#1{Appendix~{\ref{#1}}}

\def\Rcite#1{Ref.~\cite{#1}}
\def\Rcites#1{Refs.~\cite{#1}}

\newcommand*\Bell{\ensuremath{\boldsymbol\ell}}

\newcommand{\be}{\begin{equation}}
\newcommand{\ee}{\end{equation}}
\newcommand{\ba}{\begin{align}}
\newcommand{\ea}{\end{align}}
\ifx\genfrac\sdflkaj\else\fi
\newcommand{\sfrac}[2]{{\textstyle\frac{#1}{#2}}}

\begin{document}

\preprint{HU-EP-22/03-RTG}

\title{Conservative and radiative dynamics of spinning bodies at third post-Minkowskian order
using worldline quantum field theory}

\author{Gustav Uhre Jakobsen} 
\email{gustav.uhre.jakobsen@physik.hu-berlin.de}
\affiliation{%
Institut f\"ur Physik und IRIS Adlershof, Humboldt Universit\"at zu Berlin,
Zum Gro{\ss}en Windkanal 2, 12489 Berlin, Germany
}
 \affiliation{Max Planck Institute for Gravitational Physics (Albert Einstein Institute), Am M\"uhlenberg 1, 14476 Potsdam, Germany}

\author{Gustav Mogull}
\email{gustav.mogull@aei.mpg.de} 
\affiliation{%
Institut f\"ur Physik und IRIS Adlershof, Humboldt Universit\"at zu Berlin,
Zum Gro{\ss}en Windkanal 2, 12489 Berlin, Germany
}
 \affiliation{Max Planck Institute for Gravitational Physics (Albert Einstein Institute), Am M\"uhlenberg 1, 14476 Potsdam, Germany}

\begin{abstract}
Using the spinning worldline quantum field theory formalism
we calculate the quadratic-in-spin
momentum impulse $\Delta p_i^\mu$ and spin kick $\Delta a_i^\mu$
from a scattering of two arbitrarily oriented spinning massive bodies
(black holes or neutron stars) in a weak gravitational background
up to third post-Minkowskian (PM) order ($G^3$).
Two-loop Feynman integrals are performed in the potential region, yielding conservative results.
For spins aligned to the orbital angular momentum we find a conservative
scattering angle that is fully consistent with state-of-the-art post-Newtonian results.
Using the 2PM radiated angular momentum previously obtained by Plefka,
Steinhoff and the present authors we generalize the angle 
to include radiation-reaction effects,
in which case it avoids divergences in the high-energy limit.
\end{abstract}

\maketitle

Recent detections by the LIGO and Virgo collaborations
of gravitational waves emitted by binary black hole and neutron star mergers
\cite{Abbott:2016blz,LIGOScientific:2017vwq,LIGOScientific:2018mvr,LIGOScientific:2020ibl,LIGOScientific:2021usb}
have driven demand for high-precision gravitational waveform templates.
In the early stage these inspirals typically run over many cycles,
making them difficult to model using numerical techniques
\cite{Pretorius:2005gq,Campanelli:2005dd,Baker:2005vv};
yet, as the gravitational field is weak this regime is well tackled
using perturbation theory.
Often this is done in a post-Newtonian (PN)
expansion in both $G$ (Newton's constant) and $c$ (the speed of light);
however, methods involving the post-Minkowskian (PM)
expansion in $G$ are gaining prominence.

The crucial insight driving this shift is that bound orbits are closely
related to unbound scattering events,
the latter more naturally handled in the PM expansion.
A well-studied approach to the bound problem in gravity
is reverse engineering a gravitational potential from scattering data
\cite{Bjerrum-Bohr:2013bxa,Cheung:2018wkq,Neill:2013wsa,
Vaidya:2014kza,Damour:2017zjx,Bjerrum-Bohr:2019kec,Cristofoli:2020uzm},
which can in turn be used to describe bound orbits.
More recent techniques such as the Bound-to-Boundary (B2B) correspondence
directly relate bound with unbound observables
\cite{Kalin:2019rwq,Kalin:2019inp,Cho:2021arx};
scattering observables may also be used as direct input
for an effective one-body description of the bound dynamics
\cite{Buonanno:1998gg,Damour:2019lcq,Antonelli:2019ytb,Damgaard:2021rnk},
also of spinning black holes or neutron stars
\cite{Vines:2016qwa,Vines:2016unv,Vines:2017hyw,Vines:2018gqi,Bini:2017xzy,Bini:2018ywr}.

To this end an enormous effort is now underway to apply techniques
used to calculate scattering amplitudes in quantum field theory (QFT)
to the bound-state problem in gravity.
The technologies involved for both constructing integrands
and performing loop integrals are well honed
\cite{Dixon:1996wi,Elvang:2013cua,Henn:2014yza,Bern:2019prr,Weinzierl:2022eaz},
and gauge-invariant scattering observables can now be obtained directly
\cite{Kosower:2018adc,Maybee:2019jus,Cristofoli:2021vyo,Cristofoli:2021jas}
without introducing a gravitational potential.
Some impressive results have been achieved:
at 3PM (two-loop) order~\cite{Bern:2019nnu,Bern:2019crd,Bern:2020gjj,Cheung:2020gyp,DiVecchia:2020ymx}
including radiation-reaction corrections
\cite{Herrmann:2021lqe,Herrmann:2021tct,DiVecchia:2021ndb,DiVecchia:2021bdo,Heissenberg:2021tzo,Bjerrum-Bohr:2021din,Damgaard:2021ipf},
tidal effects~\cite{Cheung:2020sdj,Bern:2020uwk,AccettulliHuber:2020dal}
and most recently also at 4PM order~\cite{Bern:2021dqo,Bern:2021yeh}.
A closely related approach is heavy-particle EFT
\cite{Damgaard:2019lfh,Aoude:2020onz,Brandhuber:2021kpo,
Brandhuber:2021eyq,Aoude:2020ygw,Haddad:2021znf}.

However, QFT-based methods suffer a drawback:
the need to suppress terms that ultimately
disappear in the classical $\hbar\to0$ limit.
While the classical limit is now well understood in the non-spinning case as a soft limit
\cite{Bjerrum-Bohr:2018xdl,Kosower:2018adc,Maybee:2019jus,
Damour:2019lcq,DiVecchia:2019myk,DiVecchia:2019kta},
the situation is further complicated by the need to re-interpret
quantized spin degrees of freedom in a classical setting
\cite{Guevara:2018wpp,Bautista:2019tdr,Guevara:2019fsj,Aoude:2021oqj}.
Nevertheless, these obstacles have been successfully overcome at 2PM order
\cite{Vines:2018gqi,Bern:2020buy,Kosmopoulos:2021zoq}
up to quartic order in spin~\cite{Chen:2021qkk};
other studies of higher-spin amplitudes in this context have been done
\cite{Guevara:2017csg,Chung:2018kqs,Chung:2019duq,Arkani-Hamed:2019ymq,Guevara:2020xjx,
Bautista:2021inx,Chiodaroli:2021eug,Bautista:2021wfy,Crawley:2021auj,Guevara:2021yud,Adamo:2021rfq}.

In this regard the worldline EFT framework is more economical
\cite{Goldberger:2004jt,Goldberger:2006bd,Goldberger:2009qd,
Porto:2016pyg,Levi:2018nxp,Goldberger:2017ogt},
avoiding quantum corrections from the outset.
Partial results for the gravitational potential are now available up to 6PN order
\cite{Blumlein:2020znm,Bini:2020nsb,Bini:2020hmy,
Blumlein:2020pyo,Foffa:2020nqe,Blumlein:2021txj};
in the PM expansion recent progress has closely followed the QFT program
\cite{Kalin:2020mvi,Kalin:2020fhe,Kalin:2020lmz,Mougiakakos:2021ckm,Riva:2021vnj}
including at 4PM order~\cite{Dlapa:2021npj,Dlapa:2021vgp}.
To handle spin a local co-rotating frame is often introduced
\cite{Porto:2005ac,Levi:2015msa,Goldberger:2020fot}:
quadratic-in-spin results are available up to 5PN (N$^3$LO)
\cite{Hartung:2011ea,Levi:2011eq,Levi:2014sba,Levi:2015ixa,Levi:2016ofk,
Levi:2020kvb,Levi:2020uwu,Kim:2021rfj,Cho:2022syn}
and 2PM orders~\cite{Liu:2021zxr} --- until now at 4PN order the former
have remained unchecked.

The recently developed worldline QFT (WQFT) formalism
\cite{Mogull:2020sak,Jakobsen:2021smu,Jakobsen:2021lvp,Jakobsen:2021zvh}
innovates over these approaches by quantizing worldline degrees of freedom.
This leads to a highly streamlined PM setup wherein classical
scattering observables are directly computed as sums of tree-level Feynman diagrams.
The use of an $\cN=2$ supersymmetric extension
to the point-particle action to encapsulate spin degrees of freedom
\cite{Jakobsen:2021lvp,Jakobsen:2021zvh}
circumvents the need for a local co-rotating frame.
Recent work on the WQFT has included the double copy~\cite{Shi:2021qsb}
and applications to light bending~\cite{Bastianelli:2021nbs};
other closely related approaches involve directly solving the classical equations of motion
\cite{Saketh:2021sri} and Wilson line operators~\cite{Bonocore:2021qxh}.

In this Letter we realize the spinning WQFT's full potential
with a state-of-the-art calculation:
deriving the quadratic-in-spin conservative momentum impulse $\Delta p_i^\mu$
and spin kick $\Delta a_i^\mu$ in a scattering encounter
between massive bodies at 3PM order, including finite-size effects.
Specializing to aligned spins yields the conservative scattering angle $\theta_{\rm cons}$,
which we generalize to include dissipative effects using the
linear response relation \cite{Bini:2012ji,Damour:2020tta,Bini:2021gat}.

\sec{Spinning WQFT formalism}
The dynamics of Kerr black holes with masses $m_i$
and positions $x_i^\mu(\tau)$ on a curved $D$-dimensional background metric $g_{\mu\nu}$
are described up to quadratic order in spin by the $\cN=2$ supersymmetric worldline action
\cite{Bastianelli:2005vk,Bastianelli:2005uy}
\begin{align}\label{eq:action}
	\frac{S^{(i)}}{m_i}=-\!\int\!\d\tau \Bigl [\sfrac{1}{2}g_{\mu\nu}\dot x_i^{\mu}\dot x_i^{\nu}
	\!+\! i\bar\psi_{i,a}\!\sfrac{\D\psi_i^a}{\D\tau}\!+\!\sfrac12
	R_{abcd}\bar\psi_i^{a}\psi_i^{b}\bar\psi_i^{c}\psi_i^{d}\Bigr ].
\end{align}
The complex Grassmann-valued vectors $\psi_i^a(\tau)$,
defined in a local frame $e_\mu^a$ with $g_{\mu\nu}=e_\mu^a e_\nu^b\eta_{ab}$
and $\frac{\D\psi_i^a}{\D\tau}=\dot\psi_i^a+\dot x^\mu{\omega_\mu}^{ab}\psi_{i,b}$,
encode spin degrees of freedom (we use the mostly minus metric).
The spin tensors $S_i^{\mu\nu}$ and Pauli-Lubanski spin vectors $a_i^\mu$ are composite fields:
\begin{align}\label{eq:spinTensor}
	S_i^{\mu\nu}(\tau)=-2ie_a^\mu e_b^\nu\bar\psi_i^{[a}\psi_i^{b]}\,,\,
	a_i^\mu(\tau)=\sfrac1{2m_i}{\eps^\mu}_{\nu\rho\sigma}S_i^{\nu\rho}p_i^\sigma\,,
\end{align}
where $p_{i,\mu}=m_ig_{\mu\nu}\dot{x}_i^\nu$
(referred to as $\pi_{i,\mu}$ in \Rcite{Jakobsen:2021zvh}).

Reparametrization invariance in $\tau$ and U(1) symmetry on the Grassmann vectors
respectively imply conservation of $p_i^2$ and $\bar\psi_i\cdot\psi_i$.
Global $\cN=2$ supersymmetry provides two additional
fermionic charges: $p_i\cdot\psi_i$ and $p_i\cdot\bar\psi_i$,
which when set to zero together imply the Tulczyjew-Dixon
spin-supplementary condition (SSC) $p_{i,\mu}S_i^{\mu\nu}=0$
\cite{Tulczjew,Steinhoff:2014kwa}.
The action \eqref{eq:action}
extends naturally to include finite-size objects like neutron stars by also including
\begin{equation}\label{eq:finiteSizeDef}
  S^{(i)}_{\rm E}:=
  -m_i\, C_{{\rm E},i}\int\!\d\tau\, R_{a\mu b\nu}\dot{x}_i^\mu\dot{x}_i^\nu
  \bar\psi_i^a \psi_i^b\tilde{P}_{cd}\bar\psi_i^{c}\psi_i^{d}\,,
\end{equation}
with projector $\tilde{P}_{ab}:=\eta_{ab}-e_{a\mu}e_{b\nu}\dot{x}^\mu\dot{x}^\nu/\dot{x}^2$
and Wilson coefficients $C_{{\rm E},i}$,
where $C_{{\rm E},i}=0$ for black holes.
The projector ensures supersymmetry for terms up to $\cO(S^2)$:
enough to maintain the SSC and preserve lengths of the spin vectors.

The WQFT's distinguishing feature is quantization of both
bulk and worldline degrees of freedom.
In a weak gravitational field with $\kappa=\sqrt{32\pi G}$ we expand
$g_{\mu\nu}(x)=\eta_{\mu\nu}+\kappa h_{\mu\nu}(x)$
with the vielbein $e^{a}_{\mu}=\eta^{a\nu}\big(\eta_{\mu\nu}+ \sfrac{\kappa}{2}h_{\mu\nu} - 
\sfrac{\kappa^2}{8}h_{\mu\rho}{h^\rho}_\nu+\cdots\big)$.
Thereafter we no longer distinguish between spacetime $\mu,\nu,\ldots$
and local frame $a,b,\ldots$ indices.
The worldline fields are similarly expanded around their background values:
\begin{align}
\begin{aligned}
	x_i^\mu(\tau) &= b_i^\mu \!+\! v_i^\mu \tau \!+\! z_i^\mu(\tau)\,, &
	\psi^\mu_i(\tau) &= \Psi^\mu_i\!+\!{\psi'}_i^\mu(\tau)\,,\\
	S_i^{\mu\nu}(\tau)&={\cal S}_i^{\mu\nu}\!+\!S_i^{\prime\mu\nu}(\tau)\,, &
	a_i^\mu(\tau)&=a_{i0}^\mu\!+\!a_i^{\prime\mu}(\tau)\,,
\end{aligned}
\end{align}
where ${\cal S}_i^{\mu\nu}=-2i\bar\Psi_i^{[\mu}\Psi_i^{\nu]}$ and
$a_{i0}^\mu=\frac12{\eps^\mu}_{\nu\rho\sigma}{\cal S}_i^{\nu\rho}v_i^\sigma$.
Vanishing of the supercharges implies $v_i\cdot\Psi_i=v_i\cdot\bar\Psi=0$,
so $v_{i,\mu}\cS_i^{\mu\nu}=0$;
using $\tau$-reparametrization invariance on each worldline we fix
$b\cdot v_i=0$ where $b^\mu=b_2^\mu-b_1^\mu$.
We also define the Lorentz factor $\gamma=v_1\cdot v_2$
and the relative velocity $v=\sqrt{\gamma^2-1}/\gamma$.

The WQFT is defined by a path integral,
with physical observables calculated as operator expectation values:
\begin{align}\label{ZWQFTdef}
\begin{aligned}
	\braket{\cO}:=\!\!
	\int\!{\cal D}[h_{\mu\nu},z_i^\mu,{\psi_{i}^{\prime}}^\mu]
	e^{i (S_{\rm EH}+S_{\rm gf}+\sum_{i=1}^2S^{(i)}+S_{\rm E}^{(i)})}\cO.
\end{aligned}
\end{align}
We have included the $D$-dimensional Einstein-Hilbert action $S_{\rm EH}$
and gauge-fixing term $S_{\rm gf}$ to enforce
$\partial_\nu h^{\mu\nu}=\frac12\partial^\mu{h^\nu}_\nu$.
The stationary phase of the path integral is dominated by solutions to
the the physically relevant
Einstein and Mathisson-Papapetrou-Dixon (MPD) equations of motion
\cite{Mathisson:1937zz,Papapetrou:1951pa,Dixon:1970zza}.
This highlights the WQFT's main advantage when studying classical physics:
the classical $\hbar\to0$ limit is identified
with the sum of tree-level Feynman diagrams.

The WQFT Feynman rules are most naturally expressed in the Fourier domain:
$h_{\mu\nu}(x)=\int_k e^{-ik\cdot x}h_{\mu\nu}(k)$ and
$z_i^\mu(\tau)=\int_\omega e^{-i\omega\tau}z_i^\mu(\omega)$,
$\psi_i^{\prime\mu}(\tau)=\int_\omega e^{-i\omega\tau}\psi_i^{\prime\mu}(\omega)$,
where $\int_k:=\int\frac{\d^Dk}{(2\pi)^D}$ and $\int_\omega:=\int\frac{\d\omega}{2\pi}$.
Feynman rules for the graviton $h_{\mu\nu}$ originating from the bulk
Einstein-Hilbert action are conventional, with propagator
\begin{align}\label{eq:gravProp}
	\begin{tikzpicture}[baseline={(current bounding box.center)}]
	\coordinate (x) at (-.7,0);
	\coordinate (y) at (0.5,0);
	\draw [photon] (x) -- (y) node [midway, below] {$k$};
	\draw [fill] (x) circle (.08) node [above] {$\mu\nu$};
	\draw [fill] (y) circle (.08) node [above] {$\rho\sigma$};
	\end{tikzpicture}&=i\frac{P_{\mu\nu;\rho\sigma}}{k^2+{\rm sgn}(k^0)i0^+}
\end{align}
and $P_{\mu\nu;\rho\sigma}:=\eta_{\mu(\rho}\eta_{\sigma)\nu}-
\sfrac1{D-2}\eta_{\mu\nu}\eta_{\rho\sigma}$.
Given our current focus on conservative scattering the 
retarded $i0^+$ pole displacement here
plays no role upon integration, so we hide it in the following.
The $i0^+$ prescription is, however, significant for the worldline propagators
associated with $z_i^\mu$ and $\psi_i^{\prime\mu}$
which are, respectively
\begin{subequations}\label{eq:wlPropagators}
\begin{align}
  \begin{tikzpicture}[baseline={(current bounding box.center)}]
    \coordinate (in) at (-0.6,0);
    \coordinate (out) at (1.4,0);
    \coordinate (x) at (-.2,0);
    \coordinate (y) at (1.0,0);
    \draw [zUndirected] (x) -- (y) node [midway, below] {$\omega$};
    \draw [dotted] (in) -- (x);
    \draw [dotted] (y) -- (out);
    \draw [fill] (x) circle (.08) node [above] {$\mu$};
    \draw [fill] (y) circle (.08) node [above] {$\nu$};
  \end{tikzpicture}&=-i\frac{\eta^{\mu\nu}}{m_i(\omega+i0^+)^2}\,,\\
  \begin{tikzpicture}[baseline={(current bounding box.center)}]
    \coordinate (in) at (-0.6,0);
    \coordinate (out) at (1.4,0);
    \coordinate (x) at (-.2,0);
    \coordinate (y) at (1.0,0);
    \draw [zParticle] (x) -- (y) node [midway, below] {$\omega$};
    \draw [dotted] (in) -- (x);
    \draw [dotted] (y) -- (out);
    \draw [fill] (x) circle (.08) node [above] {$\mu$};
    \draw [fill] (y) circle (.08) node [above] {$\nu$};
  \end{tikzpicture}&=-i\frac{\eta^{\mu\nu}}{m_i(\omega+i0^+)}\,.
\end{align}
\end{subequations}
The background parameters $b_i^\mu$, $v_i^\mu$ and $\Psi_i^\mu$
(and therefore ${\cal S}_i^{\mu\nu}$, $a_{i0}^\mu$)
are identified with the far past:
$x_i^\mu(\tau)\xrightarrow{\tau\to-\infty}b_i^\mu+\tau v_i^\mu$ and
$\psi_i^\mu(\tau)\xrightarrow{\tau\to-\infty}\Psi_i^\mu$.
Expressions for Feynman vertices on the
worldline were provided in \Rcites{Jakobsen:2021lvp,Jakobsen:2021zvh}:
they are distinguished by energy conservation and carry
$m_ie^{ik\cdot b_i}\dd(k\cdot v_i+\sum_j\omega_j)$,
where $k$ is the total momentum of all outgoing gravitons and
$\omega_j$ are the energies of emitted
$z_i^\mu$, $\psi_i^{\prime\mu}$ and $\bar\psi_i^{\prime\mu}$ modes.

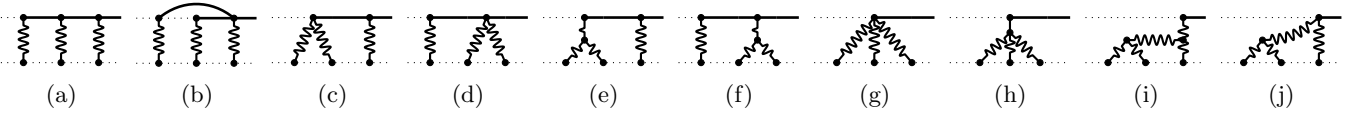
\begin{figure*}[t]
  \centering
  \begin{subfigure}{.095\textwidth}
    \centering
    \begin{tikzpicture}[baseline={(current bounding box.center)},scale=.5]
  		\coordinate (inA) at (0.4,.6);
  		\coordinate (outA) at (3.6,.6);
  		\coordinate (inB) at (0.4,-.6);
  		\coordinate (outB) at (3.6,-.6);
  		\coordinate (xA) at (1,.6);
  		\coordinate (xyA) at (1.5,.6);
  		\coordinate (yA) at (2,.6);
  		\coordinate (yzA) at (1.5,.6);
  		\coordinate (zA) at (3,.6);
  		\coordinate (xB) at (1,-.6);
  		\coordinate (yB) at (2,-.6);
  		\coordinate (zB) at (3,-.6);
  		\draw [fill] (xA) circle (.08);
  		\draw [fill] (yA) circle (.08);
  		\draw [fill] (zA) circle (.08);
  		\draw [fill] (xB) circle (.08);
  		\draw [fill] (yB) circle (.08);
  		\draw [fill] (zB) circle (.08);
  		\draw [dotted] (inA) -- (outA);
  		\draw [dotted] (inB) -- (outB);
  		\draw [zParticleF] (zA) -- (outA);
  		\draw [draw=none] (xA) to[out=40,in=140] (zA);
  		\draw [zParticleF] (xA) -- (yA);
  		\draw [zParticleF] (yA) -- (zA);
  		\draw [photon] (xA) -- (xB);
  		\draw [photon] (yA) -- (yB);
  		\draw [photon] (zA) -- (zB);
  	\end{tikzpicture}
	\caption{}
  \end{subfigure}
  \begin{subfigure}{.095\textwidth}
    \centering
    \begin{tikzpicture}[baseline={(current bounding box.center)},scale=.5]
  		\coordinate (inA) at (0.4,.6);
  		\coordinate (outA) at (3.6,.6);
  		\coordinate (inB) at (0.4,-.6);
  		\coordinate (outB) at (3.6,-.6);
  		\coordinate (xA) at (1,.6);
  		\coordinate (xyA) at (1.5,.6);
  		\coordinate (yA) at (2,.6);
  		\coordinate (yzA) at (1.5,.6);
  		\coordinate (zA) at (3,.6);
  		\coordinate (xB) at (1,-.6);
  		\coordinate (yB) at (2,-.6);
  		\coordinate (zB) at (3,-.6);
  		\draw [fill] (xA) circle (.08);
  		\draw [fill] (yA) circle (.08);
  		\draw [fill] (zA) circle (.08);
  		\draw [fill] (xB) circle (.08);
  		\draw [fill] (yB) circle (.08);
  		\draw [fill] (zB) circle (.08);
  		\draw [dotted] (inA) -- (outA);
  		\draw [dotted] (inB) -- (outB);
  		\draw [zParticleF] (zA) -- (outA);
  		\draw [zParticleF] (xA) to[out=40,in=140] (zA);
  		\draw [zParticleF] (yA) -- (zA);
  		\draw [photon] (xA) -- (xB);
  		\draw [photon] (yA) -- (yB);
  		\draw [photon] (zA) -- (zB);
  	\end{tikzpicture}
	\caption{}
  \end{subfigure}
  \begin{subfigure}{.095\textwidth}
    \centering
    \begin{tikzpicture}[baseline={(current bounding box.center)},scale=.5]
  		\coordinate (inA) at (0.4,.6);
  		\coordinate (outA) at (3.6,.6);
  		\coordinate (inB) at (0.4,-.6);
  		\coordinate (outB) at (3.6,-.6);
  		\coordinate (xA) at (1,.6);
  		\coordinate (xyA) at (1.5,.6);
  		\coordinate (yA) at (2,.6);
  		\coordinate (yzA) at (1.5,.6);
  		\coordinate (zA) at (3,.6);
  		\coordinate (xB) at (1,-.6);
  		\coordinate (yB) at (2,-.6);
  		\coordinate (zB) at (3,-.6);
  		\draw [fill] (xyA) circle (.08);
  		\draw [fill] (zA) circle (.08);
  		\draw [fill] (xB) circle (.08);
  		\draw [fill] (yB) circle (.08);
  		\draw [fill] (zB) circle (.08);
  		\draw [dotted] (inA) -- (outA);
  		\draw [dotted] (inB) -- (outB);
  		\draw [zParticleF] (zA) -- (outA);
  		\draw [draw=none] (xA) to[out=40,in=140] (zA);
  		\draw [zParticleF] (xyA) -- (yA);
  		\draw [zParticleF] (yA) -- (zA);
  		\draw [photon] (xyA) -- (xB);
  		\draw [photon] (xyA) -- (yB);
  		\draw [photon] (zA) -- (zB);
  	\end{tikzpicture}
	\caption{}
  \end{subfigure}
  \begin{subfigure}{.095\textwidth}
    \centering
    \begin{tikzpicture}[baseline={(current bounding box.center)},scale=.5]
  		\coordinate (inA) at (0.4,.6);
  		\coordinate (outA) at (3.6,.6);
  		\coordinate (inB) at (0.4,-.6);
  		\coordinate (outB) at (3.6,-.6);
  		\coordinate (xA) at (1,.6);
  		\coordinate (xyA) at (1.5,.6);
  		\coordinate (yA) at (2,.6);
  		\coordinate (yzA) at (2.5,.6);
  		\coordinate (zA) at (3,.6);
  		\coordinate (xB) at (1,-.6);
  		\coordinate (yB) at (2,-.6);
  		\coordinate (zB) at (3,-.6);
  		\draw [fill] (xA) circle (.08);
  		\draw [fill] (yzA) circle (.08);
  		\draw [fill] (xB) circle (.08);
  		\draw [fill] (yB) circle (.08);
  		\draw [fill] (zB) circle (.08);
  		\draw [dotted] (inA) -- (outA);
  		\draw [dotted] (inB) -- (outB);
  		\draw [zParticleF] (zA) -- (outA);
  		\draw [draw=none] (xA) to[out=40,in=140] (zA);
  		\draw [zParticleF] (xA) -- (yA);
  		\draw [zParticleF] (yA) -- (zA);
  		\draw [photon] (xA) -- (xB);
  		\draw [photon] (yzA) -- (yB);
  		\draw [photon] (yzA) -- (zB);
  	\end{tikzpicture}
	\caption{}
  \end{subfigure}
  \begin{subfigure}{.095\textwidth}
    \centering
    \begin{tikzpicture}[baseline={(current bounding box.center)},scale=.5]
  		\coordinate (inA) at (0.4,.6);
  		\coordinate (outA) at (3.6,.6);
  		\coordinate (inB) at (0.4,-.6);
  		\coordinate (outB) at (3.6,-.6);
  		\coordinate (xA) at (1,.6);
  		\coordinate (xyA) at (1.5,.6);
  		\coordinate (xy0) at (1.5,0);
  		\coordinate (yA) at (2,.6);
  		\coordinate (yzA) at (1.5,.6);
  		\coordinate (zA) at (3,.6);
  		\coordinate (xB) at (1,-.6);
  		\coordinate (yB) at (2,-.6);
  		\coordinate (zB) at (3,-.6);
  		\draw [fill] (xyA) circle (.08);
  		\draw [fill] (xy0) circle (.08);
  		\draw [fill] (zA) circle (.08);
  		\draw [fill] (xB) circle (.08);
  		\draw [fill] (yB) circle (.08);
  		\draw [fill] (zB) circle (.08);
  		\draw [dotted] (inA) -- (outA);
  		\draw [dotted] (inB) -- (outB);
  		\draw [zParticleF] (zA) -- (outA);
  		\draw [draw=none] (xA) to[out=40,in=140] (zA);
  		\draw [zParticleF] (xyA) -- (yA);
  		\draw [zParticleF] (yA) -- (zA);
  		\draw [photon] (xy0) -- (xyA);
  		\draw [photon] (xy0) -- (xB);
  		\draw [photon] (xy0) -- (yB);
  		\draw [photon] (zA) -- (zB);
  	\end{tikzpicture}
	\caption{}
  \end{subfigure}
  \begin{subfigure}{.095\textwidth}
    \centering
    \begin{tikzpicture}[baseline={(current bounding box.center)},scale=.5]
  		\coordinate (inA) at (0.4,.6);
  		\coordinate (outA) at (3.6,.6);
  		\coordinate (inB) at (0.4,-.6);
  		\coordinate (outB) at (3.6,-.6);
  		\coordinate (xA) at (1,.6);
  		\coordinate (xyA) at (1.5,.6);
  		\coordinate (yA) at (2,.6);
  		\coordinate (yzA) at (2.5,.6);
  		\coordinate (yz0) at (2.5,0);
  		\coordinate (zA) at (3,.6);
  		\coordinate (xB) at (1,-.6);
  		\coordinate (yB) at (2,-.6);
  		\coordinate (zB) at (3,-.6);
  		\draw [fill] (xA) circle (.08);
  		\draw [fill] (yzA) circle (.08);
  		\draw [fill] (yz0) circle (.08);
  		\draw [fill] (xB) circle (.08);
  		\draw [fill] (yB) circle (.08);
  		\draw [fill] (zB) circle (.08);
  		\draw [dotted] (inA) -- (outA);
  		\draw [dotted] (inB) -- (outB);
  		\draw [zParticleF] (zA) -- (outA);
  		\draw [draw=none] (xA) to[out=40,in=140] (zA);
  		\draw [zParticleF] (xA) -- (yA);
  		\draw [zParticleF] (yA) -- (zA);
  		\draw [photon] (xA) -- (xB);
  		\draw [photon] (yz0) -- (yzA);
  		\draw [photon] (yz0) -- (yB);
  		\draw [photon] (yz0) -- (zB);
  	\end{tikzpicture}
	\caption{}
  \end{subfigure}
  \begin{subfigure}{.095\textwidth}
    \centering
    \begin{tikzpicture}[baseline={(current bounding box.center)},scale=.5]
  		\coordinate (inA) at (0.4,.6);
  		\coordinate (outA) at (3.6,.6);
  		\coordinate (inB) at (0.4,-.6);
  		\coordinate (outB) at (3.6,-.6);
  		\coordinate (xA) at (1,.6);
  		\coordinate (xyA) at (1.5,.6);
  		\coordinate (yA) at (2,.6);
  		\coordinate (yzA) at (1.5,.6);
  		\coordinate (zA) at (3,.6);
  		\coordinate (xB) at (1,-.6);
  		\coordinate (yB) at (2,-.6);
  		\coordinate (zB) at (3,-.6);
  		\draw [fill] (yA) circle (.08);
  		\draw [fill] (xB) circle (.08);
  		\draw [fill] (yB) circle (.08);
  		\draw [fill] (zB) circle (.08);
  		\draw [dotted] (inA) -- (outA);
  		\draw [dotted] (inB) -- (outB);
  		\draw [zParticleF] (zA) -- (outA);
  		\draw [draw=none] (xA) to[out=40,in=140] (zA);
  		\draw [zParticleF] (yA) -- (zA);
  		\draw [photon] (yA) -- (xB);
  		\draw [photon] (yA) -- (yB);
  		\draw [photon] (yA) -- (zB);
  	\end{tikzpicture}
	\caption{}
  \end{subfigure}
  \begin{subfigure}{.095\textwidth}
    \centering
    \begin{tikzpicture}[baseline={(current bounding box.center)},scale=.5]
  		\coordinate (inA) at (0.4,.6);
  		\coordinate (outA) at (3.6,.6);
  		\coordinate (inB) at (0.4,-.6);
  		\coordinate (outB) at (3.6,-.6);
  		\coordinate (xA) at (1,.6);
  		\coordinate (xyA) at (1.5,.6);
  		\coordinate (yA) at (2,.6);
  		\coordinate (y0) at (2,0.2);
  		\coordinate (yzA) at (1.5,.6);
  		\coordinate (zA) at (3,.6);
  		\coordinate (xB) at (1.2,-.6);
  		\coordinate (yB) at (2,-.6);
  		\coordinate (zB) at (2.8,-.6);
  		\draw [fill] (yA) circle (.08);
  		\draw [fill] (xB) circle (.08);
  		\draw [fill] (yB) circle (.08);
  		\draw [fill] (zB) circle (.08);
  		\draw [fill] (y0) circle (.08);
  		\draw [dotted] (inA) -- (outA);
  		\draw [dotted] (inB) -- (outB);
  		\draw [zParticleF] (zA) -- (outA);
  		\draw [draw=none] (xA) to[out=40,in=140] (zA);
  		\draw [zParticleF] (yA) -- (zA);
  		\draw [photon] (y0) -- (yA);
  		\draw [photon] (y0) -- (xB);
  		\draw [photon] (y0) -- (yB);
  		\draw [photon] (y0) -- (zB);
  	\end{tikzpicture}
	\caption{}
  \end{subfigure}
  \begin{subfigure}{.095\textwidth}
    \centering
    \begin{tikzpicture}[baseline={(current bounding box.center)},scale=.5]
  		\coordinate (inA) at (0.4,.6);
  		\coordinate (outA) at (3.6,.6);
  		\coordinate (inB) at (0.4,-.6);
  		\coordinate (outB) at (3.6,-.6);
  		\coordinate (xA) at (1,.6);
  		\coordinate (xyA) at (1.5,.6);
  		\coordinate (xy0) at (1.5,0);
  		\coordinate (z0) at (3,0);
  		\coordinate (yA) at (2,.6);
  		\coordinate (yzA) at (1.5,.6);
  		\coordinate (zA) at (3,.6);
  		\coordinate (xB) at (1,-.6);
  		\coordinate (yB) at (2,-.6);
  		\coordinate (zB) at (3,-.6);
  		\draw [fill] (xy0) circle (.08);
  		\draw [fill] (zA) circle (.08);
  		\draw [fill] (xB) circle (.08);
  		\draw [fill] (yB) circle (.08);
  		\draw [fill] (zB) circle (.08);
  		\draw [fill] (z0) circle (.08);
  		\draw [dotted] (inA) -- (outA);
  		\draw [dotted] (inB) -- (outB);
  		\draw [zParticleF] (zA) -- (outA);
  		\draw [photon] (xy0) -- (z0);
  		\draw [draw=none] (xA) to[out=40,in=140] (zA);
  		\draw [photon] (xy0) -- (xB);
  		\draw [photon] (xy0) -- (yB);
  		\draw [photon] (zA) -- (zB);
  	\end{tikzpicture}
	\caption{}
  \end{subfigure}
  \begin{subfigure}{.095\textwidth}
    \centering
    \begin{tikzpicture}[baseline={(current bounding box.center)},scale=.5]
  		\coordinate (inA) at (0.4,.6);
  		\coordinate (outA) at (3.6,.6);
  		\coordinate (inB) at (0.4,-.6);
  		\coordinate (outB) at (3.6,-.6);
  		\coordinate (xA) at (1,.6);
  		\coordinate (xyA) at (1.5,.6);
  		\coordinate (xy0) at (1.5,0);
  		\coordinate (z0) at (3,0);
  		\coordinate (yA) at (2,.6);
  		\coordinate (yzA) at (1.5,.6);
  		\coordinate (zA) at (3,.6);
  		\coordinate (xB) at (1,-.6);
  		\coordinate (yB) at (2,-.6);
  		\coordinate (zB) at (3,-.6);
  		\draw [fill] (xy0) circle (.08);
  		\draw [fill] (zA) circle (.08);
  		\draw [fill] (xB) circle (.08);
  		\draw [fill] (yB) circle (.08);
  		\draw [fill] (zB) circle (.08);
  		\draw [dotted] (inA) -- (outA);
  		\draw [dotted] (inB) -- (outB);
  		\draw [zParticleF] (zA) -- (outA);
  		\draw [photon] (xy0)to[out=0,in=-140] (zA);
  		\draw [draw=none] (xA) to[out=40,in=140] (zA);
  		\draw [photon] (xy0) -- (xB);
  		\draw [photon] (xy0) -- (yB);
  		\draw [photon] (zA) -- (zB);
  	\end{tikzpicture}
	\caption{}
  \end{subfigure}
  \caption{\small
  	The ten types of diagrams contributing to the $m_1m_2^3$ components
  	of $\Delta p_1^{(3)\mu}$ and the $m_2^3$ components of $\Delta\psi_1^{(3)\mu}$,
  	involving ${\cal I}^{(1;\pm)}$-type integrals~\eqref{eq:integralFamilies}.
  	In the test-body limit $m_1\ll m_2$ these are the only surviving contributions.
  	All graphs should be considered trees ---
  	the dotted lines represent the worldlines on which energy is conserved,
  	instead of momentum.
  }
  \label{fig:testBodyDiags}
\end{figure*}

\begin{figure*}[t!]
  \centering
  \begin{subfigure}{0.12\textwidth}
    \centering
    \begin{tikzpicture}[baseline={(current bounding box.center)},scale=.6]
  		\coordinate (inA) at (0.4,.6);
  		\coordinate (outA) at (3.6,.6);
  		\coordinate (inB) at (0.4,-.6);
  		\coordinate (outB) at (3.6,-.6);
  		\coordinate (xA) at (1,.6);
  		\coordinate (xyA) at (1.5,.6);
  		\coordinate (yA) at (2,.6);
  		\coordinate (yzA) at (1.5,.6);
  		\coordinate (zA) at (3,.6);
  		\coordinate (xB) at (1,-.6);
  		\coordinate (yB) at (2,-.6);
  		\coordinate (zB) at (3,-.6);
  		\draw [fill] (xA) circle (.08);
  		\draw [fill] (yA) circle (.08);
  		\draw [fill] (zA) circle (.08);
  		\draw [fill] (xB) circle (.08);
  		\draw [fill] (yB) circle (.08);
  		\draw [fill] (zB) circle (.08);
  		\draw [dotted] (inA) -- (outA);
  		\draw [dotted] (inB) -- (outB);
  		\draw [zParticleF] (zA) -- (outA);
  		\draw [draw=none] (xA) to[out=40,in=140] (zA);
  		\draw [zParticleF] (xB) -- (yB);
  		\draw [zParticleF] (yA) -- (zA);
  		\draw [photon] (xA) -- (xB);
  		\draw [photon] (yA) -- (yB);
  		\draw [photon] (zA) -- (zB);
  	\end{tikzpicture}
	\caption{}
  \end{subfigure}
  \begin{subfigure}{0.12\textwidth}
    \centering
    \begin{tikzpicture}[baseline={(current bounding box.center)},scale=.6]
  		\coordinate (inA) at (0.4,.6);
  		\coordinate (outA) at (3.6,.6);
  		\coordinate (inB) at (0.4,-.6);
  		\coordinate (outB) at (3.6,-.6);
  		\coordinate (xA) at (1,.6);
  		\coordinate (xyA) at (1.5,.6);
  		\coordinate (yA) at (2,.6);
  		\coordinate (yzA) at (1.5,.6);
  		\coordinate (zA) at (3,.6);
  		\coordinate (xB) at (1,-.6);
  		\coordinate (yB) at (2,-.6);
  		\coordinate (zB) at (3,-.6);
  		\draw [fill] (xA) circle (.08);
  		\draw [fill] (yA) circle (.08);
  		\draw [fill] (zA) circle (.08);
  		\draw [fill] (xB) circle (.08);
  		\draw [fill] (yB) circle (.08);
  		\draw [fill] (zB) circle (.08);
  		\draw [dotted] (inA) -- (outA);
  		\draw [dotted] (inB) -- (outB);
  		\draw [zParticleF] (zA) -- (outA);
  		\draw [draw=none] (xA) to[out=40,in=140] (zA);
  		\draw [zParticleF] (xA) -- (yA);
  		\draw [zParticleF] (yB) -- (zB);
  		\draw [photon] (xA) -- (xB);
  		\draw [photon] (yA) -- (yB);
  		\draw [photon] (zA) -- (zB);
  	\end{tikzpicture}
	\caption{}
  \end{subfigure}
  \begin{subfigure}{0.12\textwidth}
    \centering
    \begin{tikzpicture}[baseline={(current bounding box.center)},scale=.6]
  		\coordinate (inA) at (0.4,.6);
  		\coordinate (outA) at (3.6,.6);
  		\coordinate (inB) at (0.4,-.6);
  		\coordinate (outB) at (3.6,-.6);
  		\coordinate (xA) at (1,.6);
  		\coordinate (xyA) at (1.5,.6);
  		\coordinate (yA) at (2,.6);
  		\coordinate (yzA) at (1.5,.6);
  		\coordinate (zA) at (3,.6);
  		\coordinate (xB) at (1,-.6);
  		\coordinate (yB) at (2,-.6);
  		\coordinate (zB) at (3,-.6);
  		\draw [fill] (xA) circle (.08);
  		\draw [fill] (yA) circle (.08);
  		\draw [fill] (zA) circle (.08);
  		\draw [fill] (xB) circle (.08);
  		\draw [fill] (yB) circle (.08);
  		\draw [fill] (zB) circle (.08);
  		\draw [dotted] (inA) -- (outA);
  		\draw [dotted] (inB) -- (outB);
  		\draw [zParticleF] (zA) -- (outA);
  		\draw [draw=none] (xA) to[out=40,in=140] (zA);
  		\draw [zParticleF] (xA) to[out=40,in=140] (zA);
  		\draw [zParticleF] (yB) -- (zB);
  		\draw [photon] (xA) -- (xB);
  		\draw [photon] (yA) -- (yB);
  		\draw [photon] (zA) -- (zB);
  	\end{tikzpicture}
	\caption{}
  \end{subfigure}
  \begin{subfigure}{0.12\textwidth}
    \centering
    \begin{tikzpicture}[baseline={(current bounding box.center)},scale=.6]
  		\coordinate (inA) at (0.4,.6);
  		\coordinate (outA) at (3.6,.6);
  		\coordinate (inB) at (0.4,-.6);
  		\coordinate (outB) at (3.6,-.6);
  		\coordinate (xA) at (1,.6);
  		\coordinate (xyB) at (1.5,-.6);
  		\coordinate (yA) at (2,.6);
  		\coordinate (yzA) at (1.5,.6);
  		\coordinate (zA) at (3,.6);
  		\coordinate (xB) at (1,-.6);
  		\coordinate (yB) at (2,-.6);
  		\coordinate (zB) at (3,-.6);
  		\draw [fill] (xA) circle (.08);
  		\draw [fill] (yA) circle (.08);
  		\draw [fill] (zA) circle (.08);
  		\draw [fill] (xyB) circle (.08);
  		\draw [fill] (zB) circle (.08);
  		\draw [dotted] (inA) -- (outA);
  		\draw [dotted] (inB) -- (outB);
  		\draw [zParticleF] (zA) -- (outA);
  		\draw [draw=none] (xA) to[out=40,in=140] (zA);
  		\draw [zParticleF] (yA) -- (zA);
  		\draw [photon] (xA) -- (xyB);
  		\draw [photon] (yA) -- (xyB);
  		\draw [photon] (zA) -- (zB);
  	\end{tikzpicture}
	\caption{}
  \end{subfigure}
  \begin{subfigure}{0.12\textwidth}
    \centering
    \begin{tikzpicture}[baseline={(current bounding box.center)},scale=.6]
  		\coordinate (inA) at (0.4,.6);
  		\coordinate (outA) at (3.6,.6);
  		\coordinate (inB) at (0.4,-.6);
  		\coordinate (outB) at (3.6,-.6);
  		\coordinate (xA) at (1,.6);
  		\coordinate (xyA) at (1.5,.6);
  		\coordinate (yA) at (2,.6);
  		\coordinate (yzB) at (2.5,-.6);
  		\coordinate (zA) at (3,.6);
  		\coordinate (xB) at (1,-.6);
  		\coordinate (yB) at (2,-.6);
  		\coordinate (zB) at (3,-.6);
  		\draw [fill] (xA) circle (.08);
  		\draw [fill] (yA) circle (.08);
  		\draw [fill] (zA) circle (.08);
  		\draw [fill] (xB) circle (.08);
  		\draw [fill] (yzB) circle (.08);
  		\draw [dotted] (inA) -- (outA);
  		\draw [dotted] (inB) -- (outB);
  		\draw [zParticleF] (zA) -- (outA);
  		\draw [draw=none] (xA) to[out=40,in=140] (zA);
  		\draw [zParticleF] (xA) -- (yA);
  		\draw [photon] (xA) -- (xB);
  		\draw [photon] (yA) -- (yzB);
  		\draw [photon] (zA) -- (yzB);
  	\end{tikzpicture}
	\caption{}
  \end{subfigure}
  \begin{subfigure}{0.12\textwidth}
    \centering
    \begin{tikzpicture}[baseline={(current bounding box.center)},scale=.6]
  		\coordinate (inA) at (0.4,.6);
  		\coordinate (outA) at (3.6,.6);
  		\coordinate (inB) at (0.4,-.6);
  		\coordinate (outB) at (3.6,-.6);
  		\coordinate (xA) at (1,.6);
  		\coordinate (xyA) at (1.5,.6);
  		\coordinate (yA) at (2,.6);
  		\coordinate (yzA) at (1.5,.6);
  		\coordinate (yzB) at (2.5,-.6);
  		\coordinate (zA) at (3,.6);
  		\coordinate (xB) at (1,-.6);
  		\coordinate (yB) at (2,-.6);
  		\coordinate (zB) at (3,-.6);
  		\draw [fill] (xA) circle (.08);
  		\draw [fill] (yA) circle (.08);
  		\draw [fill] (zA) circle (.08);
  		\draw [fill] (xB) circle (.08);
  		\draw [fill] (yzB) circle (.08);
  		\draw [dotted] (inA) -- (outA);
  		\draw [dotted] (inB) -- (outB);
  		\draw [zParticleF] (zA) -- (outA);
  		\draw [zParticleF] (xA) to[out=40,in=140] (zA);
  		\draw [photon] (xA) -- (xB);
  		\draw [photon] (yA) -- (yzB);
  		\draw [photon] (zA) -- (yzB);
  	\end{tikzpicture}
	\caption{}
  \end{subfigure}
  \begin{subfigure}{0.12\textwidth}
    \centering
    \begin{tikzpicture}[baseline={(current bounding box.center)},scale=.6]
  		\coordinate (inA) at (0.4,.6);
  		\coordinate (outA) at (3.6,.6);
  		\coordinate (inB) at (0.4,-.6);
  		\coordinate (outB) at (3.6,-.6);
  		\coordinate (xA) at (1,.6);
  		\coordinate (xyA) at (1.5,.6);
  		\coordinate (yA) at (2,.6);
  		\coordinate (yzB) at (2.5,-.6);
  		\coordinate (yz0) at (2.5,0);
  		\coordinate (zA) at (3,.6);
  		\coordinate (xB) at (1,-.6);
  		\coordinate (yB) at (2,-.6);
  		\coordinate (zB) at (3,-.6);
  		\draw [fill] (xA) circle (.08);
  		\draw [fill] (yA) circle (.08);
  		\draw [fill] (zA) circle (.08);
  		\draw [fill] (xB) circle (.08);
  		\draw [fill] (yzB) circle (.08);
  		\draw [fill] (yz0) circle (.08);
  		\draw [dotted] (inA) -- (outA);
  		\draw [dotted] (inB) -- (outB);
  		\draw [zParticleF] (zA) -- (outA);
  		\draw [draw=none] (xA) to[out=40,in=140] (zA);
  		\draw [zParticleF] (xA) -- (yA);
  		\draw [photon] (xA) -- (xB);
  		\draw [photon] (yzB) -- (yz0);
  		\draw [photon] (yA) -- (yz0);
  		\draw [photon] (zA) -- (yz0);
  	\end{tikzpicture}
	\caption{}
  \end{subfigure}
  \begin{subfigure}{0.12\textwidth}
    \centering
    \begin{tikzpicture}[baseline={(current bounding box.center)},scale=.6]
  		\coordinate (inA) at (0.4,.6);
  		\coordinate (outA) at (3.6,.6);
  		\coordinate (inB) at (0.4,-.6);
  		\coordinate (outB) at (3.6,-.6);
  		\coordinate (xA) at (1,.6);
  		\coordinate (xyA) at (1.5,.6);
  		\coordinate (yA) at (2,.6);
  		\coordinate (yzB) at (2.5,-.6);
  		\coordinate (yz0) at (2.5,0);
  		\coordinate (zA) at (3,.6);
  		\coordinate (xB) at (1,-.6);
  		\coordinate (yB) at (2,-.6);
  		\coordinate (zB) at (3,-.6);
  		\draw [fill] (xA) circle (.08);
  		\draw [fill] (yA) circle (.08);
  		\draw [fill] (zA) circle (.08);
  		\draw [fill] (xB) circle (.08);
  		\draw [fill] (yzB) circle (.08);
  		\draw [fill] (yz0) circle (.08);
  		\draw [dotted] (inA) -- (outA);
  		\draw [dotted] (inB) -- (outB);
  		\draw [zParticleF] (zA) -- (outA);
  		\draw [zParticleF] (xA) to[out=40,in=140] (zA);
  		\draw [photon] (xA) -- (xB);
  		\draw [photon] (yzB) -- (yz0);
  		\draw [photon] (yA) -- (yz0);
  		\draw [photon] (zA) -- (yz0);
  	\end{tikzpicture}
	\caption{}
  \end{subfigure}
  \begin{subfigure}{0.12\textwidth}
    \centering
    \begin{tikzpicture}[baseline={(current bounding box.center)},scale=.6]
  		\coordinate (inA) at (0.4,.6);
  		\coordinate (outA) at (3.6,.6);
  		\coordinate (inB) at (0.4,-.6);
  		\coordinate (outB) at (3.6,-.6);
  		\coordinate (xA) at (1,.6);
  		\coordinate (xyA) at (1.5,.6);
  		\coordinate (yA) at (2,.6);
  		\coordinate (yzA) at (1.5,.6);
  		\coordinate (zA) at (3,.6);
  		\coordinate (xB) at (1,-.6);
  		\coordinate (yB) at (2,-.6);
  		\coordinate (zB) at (3,-.6);
  		\draw [fill] (xyA) circle (.08);
  		\draw [fill] (zA) circle (.08);
  		\draw [fill] (xB) circle (.08);
  		\draw [fill] (yB) circle (.08);
  		\draw [fill] (zB) circle (.08);
  		\draw [dotted] (inA) -- (outA);
  		\draw [dotted] (inB) -- (outB);
  		\draw [zParticleF] (zA) -- (outA);
  		\draw [zParticleF] (yB) -- (zB);
  		\draw [photon] (xyA) -- (xB);
  		\draw [photon] (xyA) -- (yB);
  		\draw [photon] (zA) -- (zB);
  	\end{tikzpicture}
	\caption{}
  \end{subfigure}
  \begin{subfigure}{0.12\textwidth}
    \centering
    \begin{tikzpicture}[baseline={(current bounding box.center)},scale=.6]
  		\coordinate (inA) at (0.4,.6);
  		\coordinate (outA) at (3.6,.6);
  		\coordinate (inB) at (0.4,-.6);
  		\coordinate (outB) at (3.6,-.6);
  		\coordinate (xA) at (1,.6);
  		\coordinate (xyA) at (1.5,.6);
  		\coordinate (yA) at (2,.6);
  		\coordinate (yzA) at (2.5,.6);
  		\coordinate (zA) at (3,.6);
  		\coordinate (xB) at (1,-.6);
  		\coordinate (yB) at (2,-.6);
  		\coordinate (zB) at (3,-.6);
  		\draw [fill] (xA) circle (.08);
  		\draw [fill] (yzA) circle (.08);
  		\draw [fill] (xB) circle (.08);
  		\draw [fill] (yB) circle (.08);
  		\draw [fill] (zB) circle (.08);
  		\draw [dotted] (inA) -- (outA);
  		\draw [dotted] (inB) -- (outB);
  		\draw [zParticleF] (zA) -- (outA);
  		\draw [zParticleF] (xB) -- (yB);
  		\draw [zParticleF] (yzA) -- (zA);
  		\draw [photon] (xA) -- (xB);
  		\draw [photon] (yzA) -- (yB);
  		\draw [photon] (yzA) -- (zB);
  	\end{tikzpicture}
	\caption{}
  \end{subfigure}
  \begin{subfigure}{0.12\textwidth}
    \centering
    \begin{tikzpicture}[baseline={(current bounding box.center)},scale=.6]
  		\coordinate (inA) at (0.4,.6);
  		\coordinate (outA) at (3.6,.6);
  		\coordinate (inB) at (0.4,-.6);
  		\coordinate (outB) at (3.6,-.6);
  		\coordinate (xA) at (1,.6);
  		\coordinate (xyB) at (1.5,-.6);
  		\coordinate (xy0) at (1.5,0);
  		\coordinate (yA) at (2,.6);
  		\coordinate (yzA) at (1.5,.6);
  		\coordinate (zA) at (3,.6);
  		\coordinate (xB) at (1,-.6);
  		\coordinate (yB) at (2,-.6);
  		\coordinate (zB) at (3,-.6);
  		\draw [fill] (xA) circle (.08);
  		\draw [fill] (yA) circle (.08);
  		\draw [fill] (zA) circle (.08);
  		\draw [fill] (xyB) circle (.08);
  		\draw [fill] (xy0) circle (.08);
  		\draw [fill] (zB) circle (.08);
  		\draw [dotted] (inA) -- (outA);
  		\draw [dotted] (inB) -- (outB);
  		\draw [zParticleF] (zA) -- (outA);
  		\draw [zParticleF] (yA) -- (zA);
  		\draw [photon] (xA) -- (xy0);
  		\draw [photon] (yA) -- (xy0);
  		\draw [photon] (xyB) -- (xy0);
  		\draw [photon] (zA) -- (zB);
  	\end{tikzpicture}
	\caption{}
  \end{subfigure}
  \begin{subfigure}{0.12\textwidth}
    \centering
    \begin{tikzpicture}[baseline={(current bounding box.center)},scale=.6]
  		\coordinate (inA) at (0.4,.6);
  		\coordinate (outA) at (3.6,.6);
  		\coordinate (inB) at (0.4,-.6);
  		\coordinate (outB) at (3.6,-.6);
  		\coordinate (xA) at (1,.6);
  		\coordinate (xyA) at (1.5,.6);
  		\coordinate (xy0) at (1.5,0);
  		\coordinate (yA) at (2,.6);
  		\coordinate (yzA) at (1.5,.6);
  		\coordinate (zA) at (3,.6);
  		\coordinate (xB) at (1,-.6);
  		\coordinate (yB) at (2,-.6);
  		\coordinate (zB) at (3,-.6);
  		\draw [fill] (xyA) circle (.08);
  		\draw [fill] (xy0) circle (.08);
  		\draw [fill] (zA) circle (.08);
  		\draw [fill] (xB) circle (.08);
  		\draw [fill] (yB) circle (.08);
  		\draw [fill] (zB) circle (.08);
  		\draw [dotted] (inA) -- (outA);
  		\draw [dotted] (inB) -- (outB);
  		\draw [zParticleF] (zA) -- (outA);
  		\draw [zParticleF] (yB) -- (zB);
  		\draw [photon] (xy0) -- (xyA);
  		\draw [photon] (xy0) -- (xB);
  		\draw [photon] (xy0) -- (yB);
  		\draw [photon] (zA) -- (zB);
  	\end{tikzpicture}
	\caption{}
  \end{subfigure}
  \begin{subfigure}{0.12\textwidth}
    \centering
    \begin{tikzpicture}[baseline={(current bounding box.center)},scale=.6]
  		\coordinate (inA) at (0.4,.6);
  		\coordinate (outA) at (3.6,.6);
  		\coordinate (inB) at (0.4,-.6);
  		\coordinate (outB) at (3.6,-.6);
  		\coordinate (xA) at (1,.6);
  		\coordinate (xyA) at (1.5,.6);
  		\coordinate (yA) at (2,.6);
  		\coordinate (yzA) at (2.5,.6);
  		\coordinate (yz0) at (2.5,0);
  		\coordinate (zA) at (3,.6);
  		\coordinate (xB) at (1,-.6);
  		\coordinate (yB) at (2,-.6);
  		\coordinate (zB) at (3,-.6);
  		\draw [fill] (xA) circle (.08);
  		\draw [fill] (yzA) circle (.08);
  		\draw [fill] (yz0) circle (.08);
  		\draw [fill] (xB) circle (.08);
  		\draw [fill] (yB) circle (.08);
  		\draw [fill] (zB) circle (.08);
  		\draw [dotted] (inA) -- (outA);
  		\draw [dotted] (inB) -- (outB);
  		\draw [zParticleF] (zA) -- (outA);
  		\draw [zParticleF] (xB) -- (yB);
  		\draw [zParticleF] (yzA) -- (zA);
  		\draw [photon] (xA) -- (xB);
  		\draw [photon] (yz0) -- (yzA);
  		\draw [photon] (yz0) -- (yB);
  		\draw [photon] (yz0) -- (zB);
  	\end{tikzpicture}
	\caption{}
  \end{subfigure}
  \begin{subfigure}{0.12\textwidth}
    \centering
    \begin{tikzpicture}[baseline={(current bounding box.center)},scale=.6]
  		\coordinate (inA) at (0.4,.6);
  		\coordinate (outA) at (3.6,.6);
  		\coordinate (inB) at (0.4,-.6);
  		\coordinate (outB) at (3.6,-.6);
  		\coordinate (xA) at (1.2,.6);
  		\coordinate (x0) at (1.2,0);
  		\coordinate (xyA) at (1.5,.6);
  		\coordinate (yA) at (2,.6);
  		\coordinate (yzB) at (2.5,-.6);
  		\coordinate (zA) at (2.8,.6);
  		\coordinate (xB) at (1.2,-.6);
  		\coordinate (yB) at (2,-.6);
  		\coordinate (zB) at (2.8,-.6);
  		\coordinate (z0) at (2.8,0);
  		\draw [fill] (xA) circle (.08);
  		\draw [fill] (x0) circle (.08);
  		\draw [fill] (z0) circle (.08);
  		\draw [fill] (zA) circle (.08);
  		\draw [fill] (xB) circle (.08);
  		\draw [fill] (zB) circle (.08);
  		\draw [dotted] (inA) -- (outA);
  		\draw [dotted] (inB) -- (outB);
  		\draw [zParticleF] (zA) -- (outA);
  		\draw [photon] (xA) -- (xB);
  		\draw [photon] (z0) -- (x0);
  		\draw [photon] (zA) -- (zB);
  	\end{tikzpicture}
	\caption{}
  \end{subfigure}
  \begin{subfigure}{0.12\textwidth}
    \centering
    \begin{tikzpicture}[baseline={(current bounding box.center)},scale=.6]
  		\coordinate (inA) at (0.4,.6);
  		\coordinate (outA) at (3.6,.6);
  		\coordinate (inB) at (0.4,-.6);
  		\coordinate (outB) at (3.6,-.6);
  		\coordinate (xA) at (1.2,.6);
  		\coordinate (xyA) at (1.5,.6);
  		\coordinate (yA) at (2,.6);
  		\coordinate (yzB) at (2.5,-.6);
  		\coordinate (zA) at (2.8,.6);
  		\coordinate (xB) at (1.2,-.6);
  		\coordinate (yB) at (2,-.6);
  		\coordinate (zB) at (2.8,-.6);
  		\draw [fill] (xA) circle (.08);
  		\draw [fill] (zA) circle (.08);
  		\draw [fill] (xB) circle (.08);
  		\draw [fill] (zB) circle (.08);
  		\draw [dotted] (inA) -- (outA);
  		\draw [dotted] (inB) -- (outB);
  		\draw [zParticleF] (zA) -- (outA);
  		\draw [photon] (xA) -- (xB);
  		\draw [photon] (zA) -- (xB);
  		\draw [photon] (zA) -- (zB);
  	\end{tikzpicture}
	\caption{}
  \end{subfigure}
  \begin{subfigure}{0.12\textwidth}
    \centering
    \begin{tikzpicture}[baseline={(current bounding box.center)},scale=.6]
  		\coordinate (inA) at (0.4,.6);
  		\coordinate (outA) at (3.6,.6);
  		\coordinate (inB) at (0.4,-.6);
  		\coordinate (outB) at (3.6,-.6);
  		\coordinate (xA) at (1.2,.6);
  		\coordinate (xyA) at (1.5,.6);
  		\coordinate (yA) at (2,.6);
  		\coordinate (yzB) at (2.5,-.6);
  		\coordinate (zA) at (2.8,.6);
  		\coordinate (xB) at (1.2,-.6);
  		\coordinate (yB) at (2,-.6);
  		\coordinate (zB) at (2.8,-.6);
  		\draw [fill] (xA) circle (.08);
  		\draw [fill] (zA) circle (.08);
  		\draw [fill] (xB) circle (.08);
  		\draw [fill] (zB) circle (.08);
  		\draw [dotted] (inA) -- (outA);
  		\draw [dotted] (inB) -- (outB);
  		\draw [zParticleF] (zA) -- (outA);
  		\draw [photon] (xA) -- (xB);
  		\draw [photon] (xA) -- (zB);
  		\draw [photon] (zA) -- (zB);
  	\end{tikzpicture}
	\caption{}
  \end{subfigure}
  \begin{subfigure}{0.12\textwidth}
    \centering
    \begin{tikzpicture}[baseline={(current bounding box.center)},scale=.6]
  		\coordinate (inA) at (0.4,.6);
  		\coordinate (outA) at (3.6,.6);
  		\coordinate (inB) at (0.4,-.6);
  		\coordinate (outB) at (3.6,-.6);
  		\coordinate (xA) at (1.2,.6);
  		\coordinate (x0) at (1.2,0);
  		\coordinate (xyA) at (1.5,.6);
  		\coordinate (yA) at (2,.6);
  		\coordinate (yzB) at (2.5,-.6);
  		\coordinate (zA) at (2.8,.6);
  		\coordinate (xB) at (1.2,-.6);
  		\coordinate (yB) at (2,-.6);
  		\coordinate (zB) at (2.8,-.6);
  		\coordinate (z0) at (2.8,0);
  		\draw [fill] (xA) circle (.08);
  		\draw [fill] (x0) circle (.08);
  		\draw [fill] (zA) circle (.08);
  		\draw [fill] (xB) circle (.08);
  		\draw [fill] (zB) circle (.08);
  		\draw [dotted] (inA) -- (outA);
  		\draw [dotted] (inB) -- (outB);
  		\draw [zParticleF] (zA) -- (outA);
  		\draw [photon] (xA) -- (xB);
  		\draw [photon] (x0) to[out=0,in=140] (zB);
  		\draw [photon] (zA) -- (zB);
  	\end{tikzpicture}
	\caption{}
  \end{subfigure}
  \begin{subfigure}{0.12\textwidth}
    \centering
    \begin{tikzpicture}[baseline={(current bounding box.center)},scale=.6]
  		\coordinate (inA) at (0.4,.6);
  		\coordinate (outA) at (3.6,.6);
  		\coordinate (inB) at (0.4,-.6);
  		\coordinate (outB) at (3.6,-.6);
  		\coordinate (xA) at (1.2,.6);
  		\coordinate (x0) at (1.2,0);
  		\coordinate (xyA) at (1.5,.6);
  		\coordinate (yA) at (2,.6);
  		\coordinate (yzB) at (2.5,-.6);
  		\coordinate (zA) at (2.8,.6);
  		\coordinate (xB) at (1.2,-.6);
  		\coordinate (yB) at (2,-.6);
  		\coordinate (zB) at (2.8,-.6);
  		\coordinate (z0) at (2.8,0);
  		\draw [fill] (xA) circle (.08);
  		\draw [fill] (x0) circle (.08);
  		\draw [fill] (zA) circle (.08);
  		\draw [fill] (xB) circle (.08);
  		\draw [fill] (zB) circle (.08);
  		\draw [dotted] (inA) -- (outA);
  		\draw [dotted] (inB) -- (outB);
  		\draw [zParticleF] (zA) -- (outA);
  		\draw [photon] (xA) -- (xB);
  		\draw [photon] (x0) to[out=0,in=-140] (zA);
  		\draw [photon] (zA) -- (zB);
  	\end{tikzpicture}
	\caption{}
  \end{subfigure}
  \begin{subfigure}{0.12\textwidth}
    \centering
    \begin{tikzpicture}[baseline={(current bounding box.center)},scale=.6]
  		\coordinate (inA) at (0.4,.6);
  		\coordinate (outA) at (3.6,.6);
  		\coordinate (inB) at (0.4,-.6);
  		\coordinate (outB) at (3.6,-.6);
  		\coordinate (xA) at (1.2,.6);
  		\coordinate (x0) at (1.2,0);
  		\coordinate (xyA) at (1.5,.6);
  		\coordinate (yA) at (2,.6);
  		\coordinate (yzB) at (2.5,-.6);
  		\coordinate (zA) at (2.8,.6);
  		\coordinate (xB) at (1.2,-.6);
  		\coordinate (yB) at (2,-.6);
  		\coordinate (zB) at (2.8,-.6);
  		\coordinate (z0) at (2.8,0);
  		\draw [fill] (xA) circle (.08);
  		\draw [fill] (z0) circle (.08);
  		\draw [fill] (zA) circle (.08);
  		\draw [fill] (xB) circle (.08);
  		\draw [fill] (zB) circle (.08);
  		\draw [dotted] (inA) -- (outA);
  		\draw [dotted] (inB) -- (outB);
  		\draw [zParticleF] (zA) -- (outA);
  		\draw [photon] (xA) -- (xB);
  		\draw [photon] (xA) to[out=-40,in=180] (z0);
  		\draw [photon] (zA) -- (zB);
  	\end{tikzpicture}
	\caption{}
  \end{subfigure}
  \begin{subfigure}{0.12\textwidth}
    \centering
    \begin{tikzpicture}[baseline={(current bounding box.center)},scale=.6]
  		\coordinate (inA) at (0.4,.6);
  		\coordinate (outA) at (3.6,.6);
  		\coordinate (inB) at (0.4,-.6);
  		\coordinate (outB) at (3.6,-.6);
  		\coordinate (xA) at (1.2,.6);
  		\coordinate (x0) at (1.2,0);
  		\coordinate (xyA) at (1.5,.6);
  		\coordinate (yA) at (2,.6);
  		\coordinate (yzB) at (2.5,-.6);
  		\coordinate (zA) at (2.8,.6);
  		\coordinate (xB) at (1.2,-.6);
  		\coordinate (yB) at (2,-.6);
  		\coordinate (zB) at (2.8,-.6);
  		\coordinate (z0) at (2.8,0);
  		\draw [fill] (xA) circle (.08);
  		\draw [fill] (z0) circle (.08);
  		\draw [fill] (zA) circle (.08);
  		\draw [fill] (xB) circle (.08);
  		\draw [fill] (zB) circle (.08);
  		\draw [dotted] (inA) -- (outA);
  		\draw [dotted] (inB) -- (outB);
  		\draw [zParticleF] (zA) -- (outA);
  		\draw [photon] (xA) -- (xB);
  		\draw [photon] (xB) to[out=40,in=180] (z0);
  		\draw [photon] (zA) -- (zB);
  	\end{tikzpicture}
	\caption{}
  \end{subfigure}
  \begin{subfigure}{0.12\textwidth}
    \centering
    \begin{tikzpicture}[baseline={(current bounding box.center)},scale=.6]
  		\coordinate (inA) at (0.4,.6);
  		\coordinate (outA) at (3.6,.6);
  		\coordinate (inB) at (0.4,-.6);
  		\coordinate (outB) at (3.6,-.6);
  		\coordinate (xA) at (1.2,.6);
  		\coordinate (x0) at (1.2,0);
  		\coordinate (xyA) at (1.5,.6);
  		\coordinate (yA) at (2,.6);
  		\coordinate (yzB) at (2.5,-.6);
  		\coordinate (zA) at (2.8,.6);
  		\coordinate (xB) at (1.2,-.6);
  		\coordinate (yB) at (2,-.6);
  		\coordinate (zB) at (2.8,-.6);
  		\coordinate (y0) at (2,0);
  		\draw [fill] (xA) circle (.08);
  		\draw [fill] (zA) circle (.08);
  		\draw [fill] (xB) circle (.08);
  		\draw [fill] (zB) circle (.08);
  		\draw [fill] (y0) circle (.08);
  		\draw [dotted] (inA) -- (outA);
  		\draw [dotted] (inB) -- (outB);
  		\draw [zParticleF] (zA) -- (outA);
  		\draw [photon] (xA) -- (y0);
  		\draw [photon] (zA) -- (y0);
  		\draw [photon] (xB) -- (y0);
  		\draw [photon] (zB) -- (y0);
  	\end{tikzpicture}
	\caption{}
  \end{subfigure}
  \begin{subfigure}{0.12\textwidth}
    \centering
    \begin{tikzpicture}[baseline={(current bounding box.center)},scale=.6]
  		\coordinate (inA) at (0.4,.6);
  		\coordinate (outA) at (3.6,.6);
  		\coordinate (inB) at (0.4,-.6);
  		\coordinate (outB) at (3.6,-.6);
  		\coordinate (xA) at (1.2,.6);
  		\coordinate (x0) at (1.2,0);
  		\coordinate (xyA) at (1.5,.6);
  		\coordinate (yA) at (2,.6);
  		\coordinate (yzB) at (2.5,-.6);
  		\coordinate (zA) at (2.8,.6);
  		\coordinate (xB) at (1.2,-.6);
  		\coordinate (yB) at (2,-.6);
  		\coordinate (zB) at (2.8,-.6);
  		\coordinate (y0) at (2,.25);
  		\coordinate (y1) at (2,-.25);
  		\draw [fill] (xA) circle (.08);
  		\draw [fill] (zA) circle (.08);
  		\draw [fill] (xB) circle (.08);
  		\draw [fill] (zB) circle (.08);
  		\draw [fill] (y0) circle (.08);
  		\draw [fill] (y1) circle (.08);
  		\draw [dotted] (inA) -- (outA);
  		\draw [dotted] (inB) -- (outB);
  		\draw [zParticleF] (zA) -- (outA);
  		\draw [photon] (xA) -- (y0);
  		\draw [photon] (zA) -- (y0);
  		\draw [photon] (xB) -- (y1);
  		\draw [photon] (zB) -- (y1);
  		\draw [photon] (y0) -- (y1);
  	\end{tikzpicture}
	\caption{}
  \end{subfigure}
  \caption{\small
  	The 22 types of diagrams contributing to the $m_1^2m_2^2$ components of
  	$\Delta p_1^{(3)\mu}$ and the $m_1m_2^2$ components of $\Delta\psi_1^{(3)\mu}$,
  	involving ${\cal I}^{(2;\pm)}$-type integrals~\eqref{eq:integralFamilies}.
  	We exclude ``mushroom graphs'' that integrate to zero in the potential region.
  }
  \label{fig:splitDiags}
\end{figure*}
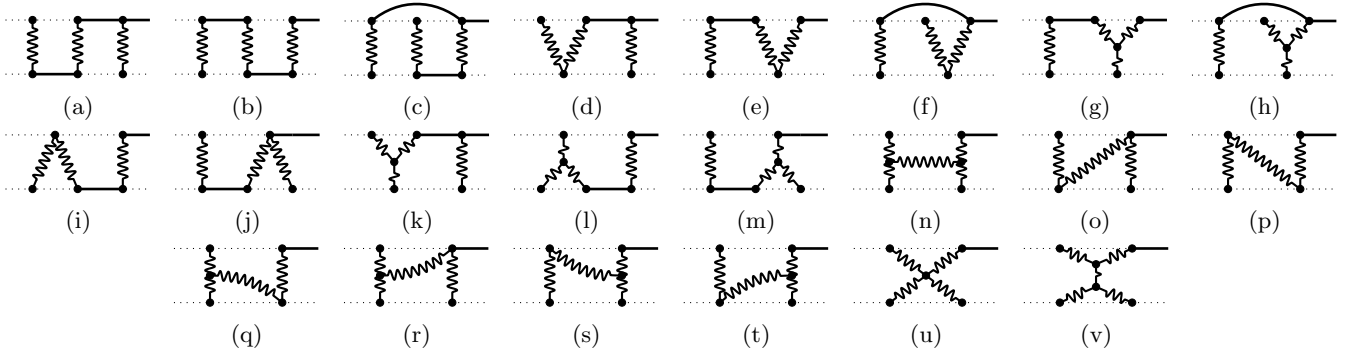

\sec{Momentum impulse and spin kick}
We derive two physical observables in an unbound scattering event:
the momentum impulse $\Delta p_i^\mu:=[p_i^\mu]^{\tau=+\infty}_{\tau=-\infty}$
and change in spin vectors $\Delta a_i^\mu:=[a_i^\mu]^{\tau=+\infty}_{\tau=-\infty}$
--- the ``spin kick''.
In the PM expansion with $\Delta X=\sum_nG^n\Delta X^{(n)}$
we focus on the 3PM components $\Delta p_1^{(3)\mu}$ and $\Delta a_1^{(3)\mu}$.
The latter we recover from both $\Delta p_1^\mu$
and $\Delta S_1^{\mu\nu}$ using \Eqn{eq:spinTensor}:
\begin{equation}
	\Delta a_i^\mu\!=\!\sfrac1{2m_i}{\eps^\mu}_{\nu\rho\sigma}\!\left(
	{\cal S}_i^{\nu\rho}\Delta p_i^\sigma\!+\!m_i\Delta S_i^{\nu\rho}v_i^\sigma\!+\!
	\Delta S_i^{\nu\rho}\Delta p_i^\sigma\right)\!,
\end{equation}
selecting the $G^3$ component on both sides.
Meanwhile $\Delta S_i^{\mu\nu}$ we derive from
$\Delta\psi_i^\mu:=[\psi_i^\mu]^{\tau=+\infty}_{\tau=-\infty}$
and $\Delta\bar\psi_i^\mu$:
\begin{equation}\label{eq:kickFromPsi}
	\Delta S_i^{\mu\nu}=
	-2i\big(\bar\Psi_i^{[\mu}\Delta\psi_i^{\nu]}+
	\Delta\bar\psi_i^{[\mu}\Psi_i^{\nu]}+
	\Delta\bar\psi_i^{[\mu}\Delta\psi_i^{\nu]}\big)\,.
\end{equation}
In the WQFT formalism these quantities are considered observables:
\begin{align}
	\Delta p_i^\mu&=
	m_i\int_{-\infty}^{\infty}\d\tau\left<\frac{\d^2x_i^\mu(\tau)}{\d\tau^2}\right>
	=-m_i\omega^2\!\left.\braket{z_i^\mu(\omega)}\right|_{\omega=0}\,,\nn \\
	\Delta\psi_i^\mu&=
	\int_{-\infty}^{\infty}\d\tau\left<\frac{\d\psi_i^\mu(\tau)}{\d\tau}\right>
	=-i\omega\!\left.\braket{\psi_i^{\prime\mu}(\omega)}\right|_{\omega=0}\,.
\end{align}
Diagrammatically this amounts to drawing
all tree-level diagrams with a single
cut external $z_i^\mu$ or $\psi_i^{\prime\mu}$ line.

The diagrams required to calculate both $\Delta p_1^{(3)\mu}$
and $\Delta\psi_1^{(3)\mu}$ are divided into three categories,
the first two of which are illustrated schematically in \Figs{fig:testBodyDiags}{fig:splitDiags}.
As the diagrams involved in $\Delta p_1^{(3)\mu}$ and $\Delta\psi_1^{(3)\mu}$ differ
only by the cut outgoing line we display them together.
For additional brevity we use only solid lines to represent propagating worldline modes
$z_i^\mu$, $\psi_i^{\prime\mu}$ and $\bar\psi_i^{\prime\mu}$;
however, it should be assumed that each internal worldline mode could be of
all three types (with expressions adjusted accordingly).
The third set of diagrams (not drawn) consists simply of 
mirrored versions of the graphs in \Fig{fig:testBodyDiags} through a horizontal plane,
but with the external cut line still on the first (upper) worldline.
For the impulse we avoid calculating these contributions directly,
instead making use of momentum conservation $\Delta p_2^{(3)\mu}=-\Delta p_1^{(3)\mu}$
(for conservative scattering).

We assemble expressions using the WQFT Feynman rules in $D=4-2\eps$ spacetime dimensions,
with the later intention of recovering four-dimensional results
in the $\eps\to0$ limit.
Each retarded graviton~\eqref{eq:gravProp} and worldline~\eqref{eq:wlPropagators} propagator
points toward the outgoing line: from cause to effect.
As diagrams belonging to each of the three categories carry common overall
factors of the masses $m_1^{\alpha}m_2^{\beta}$
the categories themselves are separately gauge invariant.
This helpfully breaks the calculation up into gauge-invariant subcomponents.
Diagrams in \Fig{fig:testBodyDiags} carry the maximum allowed power of $m_2$,
and represent the test-body limit $m_1\ll m_2$.
Integrals are performed over the energies (on the worldlines $\int_\omega$)
or momenta (in the bulk $\int_k$) of all internal lines.

The integrals involved in both $\Delta p_1^{(3)\mu}$ and $\Delta\psi_1^{(3)\mu}$
are Fourier transforms of two-loop Feynman integrals:
\begin{align}\label{eq:integralTemplate}
	\int_qe^{iq\cdot b}\dd(q\cdot v_1)\dd(q\cdot v_2)|q|^\alpha
	{\cal I}^{(i;\pm)}_{n_1,n_2,\ldots,n_7},\,\,i=1,2,3,
\end{align}
where $\dd(\omega):=2\pi\delta(\omega)$,
$q^\mu$ is the total momentum exchanged from the second to the first worldline
and $\alpha$ is an arbitrary power of $|q|:=\sqrt{-q\cdot q}$.
The two-loop integral families are
\begin{align}\label{eq:integralFamilies}
	&\!\!\!\!\!\!\!\!\!\!\!\!{\cal I}^{(1,2;\pm)}_{n_1,\ldots,n_7}
	[\ell_1^{\mu_1}\cdots\ell_1^{\mu_n}\ell_2^{\nu_1}\cdots\ell_2^{\nu_m}]\\
	:=&\int_{\ell_1,\ell_2}\!\frac{\dd(\ell_1\cdot v_2)\dd(\ell_2\cdot v_{2,1})
	\ell_1^{\mu_1}\cdots\ell_1^{\mu_n}\ell_2^{\nu_1}\cdots\ell_2^{\nu_m}}
	{D_1^{n_1}D_2^{n_2}\cdots D_7^{n_7}},\nn\\
	D_1\!=&\ell_1\!\cdot\! v_1\!+\!i0^+\,, \,
	D_2=\pm\ell_2\!\cdot\! v_{1,2}\!+\!i0^+\,,\, 
	D_3\!=\!\ell_1^2\,, \,
	D_4\!=\!\ell_2^2\,,\nn\\
	D_5\!=&(\ell_1+\ell_2-q)^2\,,\,
	D_6=(\ell_1-q)^2\,,\,
	D_7=(\ell_2-q)^2\,,\nn
\end{align}
and ${\cal I}_{n_1,\ldots,n_7}^{(3;\pm)}=
{\cal I}_{n_1,\ldots,n_7}^{(1;\pm)}\big|_{v_1\leftrightarrow v_2}$.
Each pair $(\pm)$ is associated with one of the three categories of diagrams.
To achieve these representations one must first integrate on the energies
carried by any internal deflection $z_i^\mu$
or spin $\psi_i^{\prime\mu}$, $\bar\psi_i^{\prime\mu}$ modes on the worldlines.

As two-loop integrals of this kind are now well studied ---
see e.g.~\Rcites{Bern:2019crd,Parra-Martinez:2020dzs,DiVecchia:2021bdo,Bjerrum-Bohr:2021vuf}
--- we relegate full details of how to perform them to \App{sec:Integration}.
The ${\cal I}_{n_1,\ldots,n_7}^{(1;\pm)}$ integrals ---
associated with the test-body diagrams in \Fig{fig:testBodyDiags} ---
are more straightforward, being naturally evaluated in the rest frame $v_2^\mu=(1,\mathbf{0})$.
The more involved ${\cal I}_{n_1,\ldots,n_7}^{(2;\pm)}$ integrals ---
associated with the diagrams in \Fig{fig:splitDiags} ---
contain the ${\rm arccosh}\gamma$ function.
To fix boundary conditions we adopt the \emph{potential region} of integration,
which ignores radiation-reaction contributions
and may be interpreted as a resummation of the terms
arising from a conservative PN expansion $\sfrac{v}{c}\ll1$.
We have therefore excluded certain graphs from \Fig{fig:splitDiags} ---
the so-called ``mushroom graphs'' ---
which integrate to zero within this regime.

Our final results for $\Delta p_1^{(3)\mu}$ and $\Delta a_1^{(3)\mu}$
are presented partially in \App{sec:results},
and in full in an ancillary file attached to the \texttt{arXiv} submission of this Letter.
They have the schematic form
\begin{subequations}\label{eq:schematic3PM}
\begin{align}
\Delta p_1^{(3)\mu}
  &=
  \sum_{s=0}^2
  \frac{m_1^2 m_2^2}{|b|^{3+s}}
  \Big[
  c_{0}^{(s)\mu}
  \text{arccosh}\gamma
  \!+\!
  \sum_{n=1}^3
  \Big(
  \frac{m_1}{m_2}\!
  \Big)^{n-2}
  c_{n}^{(s)\mu}
  \Big],\\
\Delta a_1^{(3)\mu}
  &=
  \sum_{s=1}^2
  \frac{m_1 m_2^2}{|b|^{2+s}}
  \Big[
  d_0^{(s)\mu}
  \text{arccosh}\gamma
  \!+\!
  \sum_{n=1}^3
  \Big(
  \frac{m_1}{m_2}\!
  \Big)^{n-2}
  d_n^{(s)\mu}
  \Big].
\end{align}
\end{subequations}
The coefficients $c_i^{(s)\mu}$ and $d_i^{(s)\mu}$ are rational functions
of $v_i^\mu$, the initial spin vectors $a_{i0}^\mu$ and the unit-normalized impact parameter
$\hat{b}^\mu:=b^\mu/|b|$, where $|b|:=\sqrt{-b\cdot b}$.
We have performed several consistency checks.
Firstly, all poles in $\eps=2-\frac{D}2$ arising from the
dimensionally regularized two-loop integrals \eqref{eq:integralFamilies} are seen to cancel,
thus ensuring finiteness of our results in the limit $D\to4$.
Secondly, conservation of $p_i^2$, $\bar\psi_i\cdot\psi_i$
and the fermionic supercharge $p_i\cdot\psi_i$
between initial and final states implies a set of consistency requirements:
\begin{align}\label{eq:checks}
	&0=m_1v_1\!\cdot\!\Delta p_1^{(3)}\!+\!\Delta p_{1}^{(1)}\!\cdot\!\Delta p_1^{(2)}\,,\\
	&0=\bar\Psi_1\!\cdot\!\Delta\psi_1^{(3)}\!+\!\Delta\bar\psi_1^{(3)}\!\cdot\!\Psi_1\!+\!
	\Delta\bar\psi_1^{(1)}\!\cdot\!\Delta\psi_1^{(2)}\!+\!
	\Delta\bar\psi_1^{(2)}\!\cdot\!\Delta\psi_1^{(1)},\nn\\
	&0=m_1v_1\!\cdot\!\Delta\psi_1^{(3)}\!+\!\Delta p^{(3)}_1\!\cdot\!\Psi_1\!+\!
	\Delta p_1^{(1)}\!\cdot\!\Delta\psi_1^{(2)}\!+\!
	\Delta p_1^{(2)}\!\cdot\!\Delta\psi_1^{(1)}\!\!.\nn
\end{align}
All three of these checks are highly nontrivial:
for instance, the third compares parts of $\Delta\psi_1^{(3)\mu}$ containing
${\rm arccosh}\gamma$ with $\Delta p_1^{(3)\mu}$ at different orders in spin.

\sec{Scattering angle}
We now specialize to spin vectors $a_i^\mu$
aligned with the orbital angular momentum: $a_i^\mu=s_il^\mu$, where
$l^\mu:={\eps^\mu}_{\nu\rho\sigma}\hat{b}^\nu v_1^\rho v_2^\sigma/(\gamma v)$,
confining the motion to a plane.
The conservative part of the scattering angle is then given by
(see e.g.~\Rcite{Kalin:2020mvi}):
\begin{equation}
	\sin\!\left(\frac{\theta_{\rm cons}}2\right)=\frac{|\Delta p_1|}{2p_\infty}\,.
\end{equation}
with the full scattering angle (including radiative corrections)
given by $\theta=\theta_{\rm cons}+\theta_{\rm rad}$.
The center-of-mass momentum is $p_\infty=\mu\gamma v/\Gamma$
where $\mu=M\nu=m_1 m_2/M$ is the symmetric mass,
$M=m_1+m_2$ is the total mass and
$\Gamma=E/M=\sqrt{1+2\nu(\gamma-1)}$,
$E$ being the total energy.
We decompose the scattering angle as
\begin{align}
  \frac{\theta}{\Gamma}&=
  \sum_n
  \bigg(
  \frac{GM}{|b|}
  \bigg)^n
  \theta^{(n)}\,,&
  \theta^{(n)}&=
  \sum_m
  \frac{\theta^{(n,m)}}{|b|^m}\,,
\end{align}
with $n$ and $m$ counting the PM and spin orders respectively.
At 3PM order using our results
\begin{subequations}\label{eq:angleCons}
\begin{widetext}
  \begin{align}
    &
  \theta^{(3,0)}_{\rm cons}
  =
  2
  \frac{
    64\gamma^6-120\gamma^4+60\gamma^2-5
  }{
    3(\gamma^2-1)^3
  }
  \Gamma^2
  -
  \frac{8\nu
  \gamma
  (14\gamma^2+25)}{3(\gamma^2-1)}
  -
  8\nu
  \frac{
    4\gamma^4-12\gamma^2-3
  }{
    (\gamma^2-1)^{3/2}
  }
  \text{arccosh}\gamma
  \ ,
  \\
  &
  \theta^{(3,1)}_{\rm cons}
  =
  2\gamma
  \frac{
    16\gamma^4-20\gamma^2+5
  }{
    (\gamma^2-1)^{5/2}
  }
  (5\Gamma^2 s_+ - \delta s_-)
  -
  4\nu s_+
  \bigg(
  \frac{44\gamma^4+100\gamma^2+41}{(\gamma^2-1)^{3/2}}
  +12\gamma\frac{(\gamma^2-6)(2\gamma^2+1)}{(\gamma^2-1)^2}
  \text{arccosh}\gamma
  \bigg)
  \ ,
  \\
  &
  \theta^{(3,2)}_{\rm cons}
  =
  \frac{4\Gamma^2}{(\gamma^2-1)^3}
  \bigg(
  (96\gamma^6\!-\!160\gamma^4\!\!+70\gamma^2\!-\!5)s_+^2
  -
  \frac{1772\gamma^6-2946\gamma^4+1346\gamma^2-137}{35} s_{{\rm E},+}^2
  \bigg)
  -
  8\delta\bigg(
  \frac{16\gamma^4-12\gamma^2+1}{(\gamma^2-1)^2}s_- s_+
  \nonumber\\&\quad
  -
  \frac{214\gamma^4-223\gamma^2+44}{35(\gamma^2-1)^2} s_{{\rm E},-}^2
  \bigg)
    +
    8\nu\gamma
    \bigg[
     \frac{2\gamma^4+86\gamma^2+87}{5(\gamma^2-1)^2}
    s_-^2
    -
    \frac{298\gamma^4+834\gamma^2+853}{5(\gamma^2-1)^2}
    s_+^2
    +
    \frac{3244\gamma^4+7972\gamma^2+4639}{105(\gamma^2-1)^2}
    s_{{\rm E},+}^2
    \nonumber\\&\quad
    -
    \bigg(
    3s_-^2
    (4\gamma^4+7\gamma^2+1)
    +
    3s_+^2
    (8\gamma^6-68\gamma^4-63\gamma^2-9)
    -
    2s_{{\rm E},+}^2
    (8\gamma^6-56\gamma^4-24\gamma^2-3)
    \bigg)
    \frac{\text{arccosh}\gamma}{\gamma(\gamma^2-1)^{5/2}}
    \bigg]
    \ ,
\end{align}
\end{widetext}
\end{subequations}
where we have defined $\delta=(m_2-m_1)/M$
as well as $s_\pm=s_1\pm s_2$ and $s_{{\rm E},\pm}^2=C_{{\rm E},1}s_1^2\pm C_{{\rm E},2}s_2^2$.
We have checked $\theta^{(3)}_{\rm cons}$ both in the test-body limit $\nu\to0$
and up to 4PN order (N$^2$LO) for comparable masses
against~\Rcites{Vines:2018gqi,Antonelli:2020ybz}.\footnote{We
	thank Mohammed Khalil for providing us with an extension of the 4PN
	scattering angle to include finite-size $C_{{\rm E},i}$ coefficients.
}
For aligned spins, this provides a first
check on the complete quadratic-in-spin conservative dynamics
of compact binaries at 4PN order~\cite{Levi:2015ixa,Levi:2016ofk}
together with recent work in the worldline EFT formalism~\cite{Cho:2022syn}.

As explained by Bini and Damour \cite{Bini:2012ji,Damour:2020tta,Bini:2021gat},
the conservative scattering angle is generalized to include
radiation using the linear response relation:
\begin{align}\label{eq:linResponse}
  \theta_{\rm rad}
  =
  -
  \frac12
  \frac{\partial\theta_{\rm cons}}{\partial E}E_{\rm rad}
  -
  \frac12
  \frac{\partial\theta_{\rm cons}}{\partial J}J_{\rm rad}
  \,.  
\end{align}
Here $J$ is the total angular momentum in the center-of-mass frame:
the derivative is equivalent to one with respect to the 
orbital angular momentum $L=p_\infty |b|$.
It has recently been clarified \cite{Veneziano:2022zwh} that
Eq.~\eqref{eq:linResponse} applies only using an ``intrinsic''
gauge choice  with respect to BMS symmetry,
wherein the radiated angular momentum $J_{\rm rad}$ begins at $\cO(G^2)$.
With $E_{\rm rad}$ starting at $\cO(G^3)$
to deduce $\theta^{(3)}_{\rm rad}$ we
need only $J^{(2)}_{\rm rad}$,
which was provided by Plefka, Steinhoff and the present authors
for arbitrary spin orientations in Ref.~\cite{Jakobsen:2021lvp}.
For aligned spins, 
\begin{align}\label{eq:radAngMom}
  &\frac{J^{(2)}_{\rm rad}}{L}=
  \bigg(
  1+\frac{2vs_+}{|b|(1+v^2)}
  +\frac{
    s_+^2
    -
    s_{{\rm E},+}^2
  }{|b|^2}
  \bigg)
  \\
  &
  \times
  \frac{4m_1m_2}{|b|^2}\frac{(2\gamma^2-1)}{\sqrt{\gamma^2-1}}
  \left(-\frac83+\frac1{v^2}+\frac{(3v^2-1)}{v^3}{\rm arccosh}\gamma\right)\,.\nn
\end{align}
This yields the radiative part of the scattering angle:
\begin{align}\label{eq:angleRad}
  &
  \theta_{\rm rad}^{(3)}
  =
  \frac{4\nu(2\gamma^2-1)^2}{(\gamma^2-1)^{3/2}}
  \Big(
  -\frac83+\frac1{v^2}+\frac{(3v^2-1)}{v^3}{\rm arccosh}\gamma\Big)
  \nn\\&\times
  \bigg[
  1+\frac{6\gamma^2v}{(2\gamma^2-1)}\frac{s_+}{|b|}
  +4
  \bigg(
    \frac{6\gamma^4-6\gamma^2+1}{(2\gamma^2-1)^2}
  \frac{s_+^2}{|b|^2}
  -
  \frac{s_{{\rm E},+}^2}{|b|^2}
  \bigg)
  \bigg]\,.
\end{align}
The non-spinning part of $\theta_{\rm rad}^{(3)}$ has also been
confirmed without reference to Eq.~\eqref{eq:linResponse} ---
see e.g.~Refs.~\cite{Herrmann:2021lqe,Herrmann:2021tct}.

A key criterion of $\theta_{\rm rad}^{(3)}$ is that
the total scattering angle should remain finite in the high-energy limit.
We write $\theta(E,\nu,|b|,\gamma,s_i)$
in terms of the energy, symmetric mass ratio, impact parameter,
Lorentz factor and spin magnitudes and let $\gamma\to\infty$,
in which case the individual masses are negligible.
In this limit:
\begin{align}
  &\theta=
  4
  \frac{G E}{|b|}
  \bigg(\!
  1\!+\!\frac{s_+}{|b|}\!+\!\frac{s_+^2-s_{{\rm E},+}^2}{|b|^2}
  \bigg)
  +
  \frac{32}{3}
  \bigg(
  \frac{GE}{|b|}
  \bigg)^3
  \bigg[
  1
  \!+\!
  3\frac{s_+}{|b|}
  \nn\\&\qquad
  +
  \frac{3}{20}\frac{41s_+^2+s_-^2-16 s_{{\rm E},+}^2}{|b|^2}
  \bigg]
  +\mathcal{O}(G^4,\gamma^{-1/2})\,.
\end{align}
While we know of no spinning extension to 
Amati, Ciafaloni and Veneziano's result~\cite{Amati:1990xe}
to compare with in the high-energy limit,
we do see that a logarithmic divergence appearing in the conservative
part of the angle \eqref{eq:angleCons} is canceled by
the radiative correction \eqref{eq:angleRad}.

\sec{Discussion}
We conclude with a brief discussion of bound observables.
Using the B2B dictionary \cite{Kalin:2019rwq,Kalin:2019inp,Cho:2021arx}
one may, for instance, recover the aligned-spin periastron advance $\Delta\Phi$
from our scattering angle:
\begin{equation}\label{eq:B2B}
	\Delta\Phi=\theta(E,L,m_i,s_i)+\theta(E,-L,m_i,-s_i)\,.
\end{equation}
Similarly one may relate the unbound and bound radial actions,
from which the scattering angle and periastron advance are
respectively given by a derivative with respect to $L$.
At 3PM order $\theta^{(3)}$ cancels in \Eqn{eq:B2B};
nevertheless, from $\theta^{(3)}$ one may reconstruct the leading-PN parts
of $\theta^{(4)}$ and $\theta^{(6)}$
(and similarly for the radial action)~\cite{Antonelli:2020ybz}.
This suffices for a comparison with bound quadratic-in-spin results at N$^2$LO:
for example, we have reproduced the quadratic-in-spin N$^2$LO binding energy
for circular orbits \cite{Levi:2015ixa,Levi:2016ofk}
as was also very recently done in~\Rcite{Cho:2022syn}.

For arbitrarily aligned spins there is currently no extension of the B2B map~\eqref{eq:B2B}.
An alternative would therefore be to make an ansatz for a conservative two-body Hamiltonian ---
for example, building on that used at 2PM order \cite{Bern:2020buy,Kosmopoulos:2021zoq} ---
and solve Hamilton's equations for comparison with $\Delta p_1^\mu$ and $\Delta a_1^\mu$, thus extending those results to 3PM.
On the other hand, we are hopeful that direct maps between
unbound and bound gauge-invariant observables for arbitrary spins
will be discovered in the near future.
In that spirit, all information is captured by the impulse and spin kick.

There remains much work to be done:
for example, extending $\Delta p_1^{(3)\mu}$ and $\Delta a_1^{(3)\mu}$
to incorporate radiation-reaction effects,
as we have already done for the scattering angle
$\theta^{(3)}_{\rm rad}$ \eqref{eq:angleRad}.
This requires us to upgrade our two-loop master integrals to account
for the retarded pole displacement on the
graviton propagator~\eqref{eq:gravProp}
and restore the mushroom graphs to Fig.~\ref{fig:splitDiags}.
We are also interested in the eikonal phase, which was computed in
\Rcite{Jakobsen:2021zvh}  at 2PM order as the free energy of the WQFT,
and captures both the impulse and spin kick.
Nevertheless, for the time being we believe that we have effectively
showcased the spinning WQFT's utility and efficiency.

\sec{Acknowledgments}
We thank Alessandra Buonanno, Jan Plefka, Muddu Saketh, Benjamin Sauer,
Jan Steinhoff and especially Justin Vines for enlightening discussions.
We also thank Gregor K\"alin, Zhengwen Liu and Rafael Porto for sharing expressions for some of the two-loop
master integrals with us (with which we agree)
and Alexander Broll for performing numerical checks.
This work is funded by the Deutsche Forschungsgemeinschaft
(DFG, German Research Foundation)
Projekt-nummer 417533893/GRK2575 ``Rethinking Quantum Field Theory''.

\newpage

\begin{widetext}

\appendix

\section{Two-loop integration}
\label{sec:Integration}

In this Appendix we outline the steps required to perform
integrals of the kind appearing in \Eqn{eq:integralTemplate}.
All scalar two-loop integrals ${\cal I}_{n_1,\ldots,n_7}^{(i;\pm)}$
(those without free indices)
are functions only of $|q|$, $\gamma$ and the dimensional
regularization parameter $\eps=2-\sfrac{D}2$.
The $|q|$ dependence is easily established as an overall factor,
$|q|$ being the only dimensionful scale.
The ${\cal I}_{n_1,\ldots,n_7}^{(1;\pm)}$ integrals
further factorize into functions of $\gamma$ and $\eps$:
one sees this by working in the rest frame of the second body $v_2^\mu=(1,\mathbf{0})$,
wherein the two $\dd$-functions are naturally resolved as $\ell_1^0=\ell_2^0=0$.

\sec{1.~Tensor reduction}
Two-loop tensor integrals ${\cal I}^{(i;\pm)}_{n_1,\ldots,n_7}
[\ell_1^{\mu_1}\cdots\ell_1^{\mu_n}\ell_2^{\nu_1}\cdots\ell_2^{\nu_m}]$
in arbitrary $D$ dimensions are decomposed onto bases consisting of
the vectors $v_i^\mu$ and $q^\mu$ appearing in their expressions, plus the metric.
We find it convenient to introduce
\begin{align}
	w^\mu_1&=\frac{\gamma v_2^\mu-v_1^\mu}{\gamma^2-1}\,, &
	w^\mu_2&=\frac{\gamma v_1^\mu-v_2^\mu}{\gamma^2-1}\,, &
	P^{\mu\nu}&=\eta^{\mu\nu}-w_1^{\mu}v_1^{\nu}-w_2^{\mu}v_2^{\nu}
	+\frac{q^\mu q^\nu}{|q|^2}\,,
\end{align}
where by design dual vectors $w_i^\mu$ satisfy $v_i\cdot w_j=\delta_{ij}$;
$P^{\mu\nu}$ is the metric of the $(D-3)$-dimensional space
orthogonal to $v_i^\mu$ and $q^\mu$.
Then, for example
\begin{subequations}
\begin{align}
	{\cal I}^{(i;\pm)}_{n_1,\ldots,n_7}[\ell_1^\mu]&=
	{\cal I}^{(i;\pm)}_{n_1,\ldots,n_7}[\ell_1\cdot v_1]w_1^\mu-
	|q|^{-2}{\cal I}^{(i;\pm)}_{n_1,\ldots,n_7}[\ell_1\cdot q]q^\mu\,,\\
	{\cal I}^{(i;\pm)}_{n_1,\ldots,n_7}[\ell_1^\mu\ell_1^\nu]&=
	\sfrac1{D-3}{\cal I}^{(i;\pm)}_{n_1,\ldots,n_7}[\ell_1\cdot P\cdot\ell_1]P^{\mu\nu}+
	{\cal I}^{(i;\pm)}_{n_1,\ldots,n_7}[(\ell_1\cdot v_1)^2]w_1^\mu w_1^\nu
	-|q|^{-2}
	{\cal I}^{(i;\pm)}_{n_1,\ldots,n_7}[\ell_1\cdot v_1\ell_1\cdot q]w_1^{(\mu}q^{\nu)}\\
	&\qquad
	+|q|^{-4}{\cal I}^{(i;\pm)}_{n_1,\ldots,n_7}[(\ell_1\cdot q)^2]q^\mu q^\nu\,.\nn
\end{align}
\end{subequations}
The resulting integrals are straightforwardly reduced to scalar-type:
\begin{align}
{\cal I}^{(i;\pm)}_{n_1,n_2,\ldots,n_7}[\ell_1\cdot q]=
\sfrac12\big({\cal I}^{(i;\pm)}_{n_1,n_2,n_3-1,n_4,n_5,n_6,n_7}
-{\cal I}^{(i;\pm)}_{n_1,n_2,n_3,n_4,n_5,n_6-1,n_7}-
|q|^2{\cal I}^{(i;\pm)}_{n_1,n_2,n_3,n_4,n_5,n_6,n_7}\big)\,.
\end{align}
The highest-rank integrals we encountered in this project had five free indices:
${\cal I}^{(i;\pm)}_{n_1,n_2,\ldots,n_7}
[\ell_1^{\mu_1}\ell_1^{\mu_2}\ell_1^{\mu_3}\ell_2^{\nu_1}\ell_2^{\nu_2}]$.

\sec{2.~Integration-by-parts relations}
Our next task is to find linear identities satisfied by the scalar integrals,
and thus establish a minimal basis --- the so-called \emph{master integrals}.
We do this separately for each of the
six integral families ${\cal I}_{n_1,\ldots,n_7}^{(i;\pm)}$.
These linear identities are generated by integration-by-parts relations (IBPs):
\begin{align}
	0=\int_{\ell_1,\ell_2}\frac{\partial}{\partial\ell^\mu_i}
	\left[k_j^\mu\frac{\dd^{(n_8)}(\ell_1\cdot v_2)\dd^{(n_9)}(\ell_2\cdot v_{2,1})}
	{(\ell_1\!\cdot\! v_1\!+\!i0^+)^{n_1}
	(\pm\ell_2\!\cdot\! v_{1,2}\!+\!i0^+)^{n_2}(\ell_1^2)^{n_3}(\ell_2^2)^{n_4}
	((\ell_1+\ell_2-q)^2)^{n_5}((\ell_1-q)^2)^{n_6}
	((\ell_2-q)^2)^{n_7}}\right]\,,
\end{align}
where $k_j^\mu=\{q^\mu,v_1^\mu,v_2^\mu,\ell_1^\mu,\ell_2^\mu\}$.
Following the \emph{reverse unitarity} approach of \Rcites{Herrmann:2021lqe,Herrmann:2021tct}
we have generalized the definition of the $\dd$-functions to include derivatives:
\begin{equation}
	\frac{\dd^{(n)}(\omega)}{(-1)^nn!}=
	\frac{i}{(\omega+i0^+)^{n+1}}-\frac{i}{(\omega-i0^+)^{n+1}}\,.
\end{equation}
From the perspective of IBPs the $\dd$-functions may thus also be regarded as (cut) propagators,
so here the integral families have nine propagators instead of seven.
The IBPs are linear relationships of the form
\begin{align}
	0=\sum_j\alpha_j{\cal I}^{(i;\pm)}_{n_1+b_{j,1},n_2+b_{j,2},\ldots,n_9+b_{j,9}}\,,
\end{align}
where $b_{j,k}=\{-1,0,1\}$ on a case-by-case basis
and $\alpha_j$ are generically functions of $\gamma$, $|q|$ and $\eps=2-\sfrac{D}2$.
The two new indices $n_8$, $n_9$ represent
the propagators associated with the delta functions ---
in general, we always choose masters with $n_8=n_9=0$.
The IBPs separate into two categories
depending on whether $n_1+n_2$ is even or odd.
The integral families also enjoy symmetry relations:
\begin{align}
	{\cal I}^{(i;\pm)}_{n_1,n_2,n_3,n_4,n_5,n_6,n_7}&=
	{\cal I}^{(i;\pm)}_{n_2,n_1,n_4,n_3,n_5,n_7,n_6}\,, &
	{\cal I}^{(i;\pm)}_{n_1,n_2,n_3,n_4,n_5,n_6,n_7}&=
	{\cal I}^{(i;\pm)}_{n_1,n_2,n_6,n_7,n_5,n_3,n_4}\,,
\end{align}
where the first is due to $\ell_1\leftrightarrow\pm\ell_2$ symmetries
and the second due to shifts in $\ell_i$ by $q$.
Using the Laporta algorithm~\cite{Laporta:1996mq,Laporta:2000dsw},
implemented in publicly-available packages such as
\texttt{FIRE}~\cite{Smirnov:2019qkx}, \texttt{LiteRed}~\cite{Lee:2012cn,Lee:2013mka}
and \texttt{KIRA}~\cite{Maierhofer:2017gsa,Klappert:2020nbg},
we reduce to minimal bases:
two bases for each family, with even or odd $n_1+n_2$.

\sec{3.~Insertion of master integrals}
We now provide expressions for the master integrals,
beginning with the simpler ${\cal I}^{(1;\pm)}$ families
to all orders in $\eps=2-\sfrac{D}2$:
\begin{subequations}\label{eq:easyMasters}
\begin{align}
	{\cal I}^{(1;\pm)}_{0,0,1,1,1,0,0}&=
	-(4\pi)^{-3+2\eps}\frac{\Gamma^3(\frac12-\eps)\Gamma(2\eps)}{\Gamma(\frac32-3\eps)}\,,\\
	{\cal I}^{(1;\pm)}_{1,0,1,1,1,0,0}&=
	(4\pi)^{-\frac52+2\eps}
	\frac{i}{2\sqrt{\gamma^2-1}}
	\frac{\Gamma(\frac12-2\eps)\Gamma^2(\frac12-\eps)\Gamma(-\eps)\Gamma(\frac12+2\eps)}
	{\Gamma(\frac12-3\eps)\Gamma(1-2\eps)}\,,\\
	{\cal I}^{(1;+)}_{1,1,1,1,1,0,0}&=
	2{\cal I}^{(1;-)}_{1,1,1,1,1,0,0}
	=(4\pi)^{-2+2\eps}\frac{\Gamma^3(-\eps)\Gamma(1+2\eps)}{3(\gamma^2-1)\Gamma(-3\eps)}\,,
\end{align}
\end{subequations}
where we have set $|q|=1$.
As promised the dependence on $\gamma$ and $\eps$ factorizes.
These results are well-established,
and may be found in e.g.~\Rcite{Parra-Martinez:2020dzs}.
The ${\cal I}^{(2;\pm)}$ master integrals do not factorize,
and we provide them only up to the order in $\eps$ to which they are required.
For $n_1+n_2$ even:
\allowdisplaybreaks
\begin{subequations}
\begin{align}
	{\cal I}^{(2;\pm)}_{0,0,0,1,1,0,1}&=0\,,\\
	{\cal I}^{(2;\pm)}_{0,0,1,1,0,1,1}&=(4\pi)^{-3+2\eps}
	\frac{\Gamma^4(\frac12-\eps)\Gamma^2(\frac12+\eps)}{\Gamma^2(1-2\eps)}
	\label{eq:prodIntegral}\,,\\
	{\cal I}^{(2;\pm)}_{0,0,1,1,1,0,0}&=-(4\pi)^{-2+2\eps}e^{-2\eps\gamma_{\rm E}}
	\frac{{\rm arccosh}\gamma}{4\eps\sqrt{\gamma^2-1}}+\cO(\eps^0)\,,\\
	{\cal I}^{(2;\pm)}_{0,0,2,1,1,0,0}&=-(4\pi)^{-2+2\eps}e^{-2\eps\gamma_{\rm E}}
	\frac{(1-2\eps)\gamma\sqrt{\gamma^2-1}+2\eps(\gamma^2-1){\rm arccosh}\gamma}
	{2\sqrt{\gamma^2-1}}+\cO(\eps^2)\,,\\
	{\cal I}^{(2;\pm)}_{0,0,1,1,2,0,0}&=-(4\pi)^{-2+2\eps}e^{-2\eps\gamma_{\rm E}}
	\frac{{\rm arccosh}\gamma}{2\sqrt{\gamma^2-1}}+\cO(\eps)\,,\\
	{\cal I}^{(2;\pm)}_{0,0,1,1,1,1,1}&=(4\pi)^{-2+2\eps}e^{-2\eps\gamma_{\rm E}}
	\frac{{\rm arccosh}\gamma+\eps({\rm arccosh}^2\gamma+{\rm Li}_2)}
	{2\eps\sqrt{\gamma^2-1}}+\cO(\eps)
	\label{eq:dilog1}\,,\\
	{\cal I}^{(2;\pm)}_{0,0,1,1,2,1,1}&=(4\pi)^{-2+2\eps}e^{-2\eps\gamma_{\rm E}}
	\frac{(1+5\eps)\gamma\sqrt{\gamma^2-1}-(1+\eps+2\gamma^2\eps){\rm arccosh}\gamma
	-\eps({\rm arccosh}^2\gamma+{\rm Li}_2)}
	{2\sqrt{\gamma^2-1}}+\cO(\eps^2)
	\label{eq:dilog2}\,,\\
	{\cal I}^{(2;+)}_{1,1,1,1,1,0,0}&=
	\frac12{\cal I}^{(2;-)}_{1,1,1,1,1,0,0}=
	(4\pi)^{-2+2\eps}e^{-2\eps\gamma_{\rm E}}
	\frac{1}{2\eps^2(\gamma^2-1)}+\cO(\eps^{-1})\,,
\end{align}
\end{subequations}
where the first two are known to all orders in $\eps$ ---
the integral \eqref{eq:prodIntegral} is a product of one-loop integrals.
The dilogarithm appearing in the integrals \eqref{eq:dilog1} and \eqref{eq:dilog2}
is ${\rm Li}_2(2-2\gamma^2+2\gamma\sqrt{\gamma^2-1})$:
this dilogarithm and ${\rm arccosh}^2\gamma$ cancel from all of our final results
between these two integrals.
In the non-spinning part of $\Delta p_1^{(3)\mu}$~\eqref{eq:nonSpinImpulse}
these eight master integrals are associated with
terms proportional to the impact parameter $b^\mu$.
For $n_1+n_2$ odd:
\begin{subequations}
\begin{align}
	{\cal I}^{(2;\pm)}_{1,0,1,0,1,1,0}&=0\,,\\
	{\cal I}^{(2;\pm)}_{1,0,0,1,1,0,1}&=
	(2\pi)^{-1+2\eps}e^{-2\eps\gamma_{\rm E}}\frac{i}{32\eps\sqrt{\gamma^2-1}}
	+\cO(\eps^0)\,,\\
	{\cal I}^{(2;\pm)}_{1,0,1,1,1,0,0}&=
	-(2\pi)^{-1+2\eps}e^{-2\eps\gamma_{\rm E}}\frac{i}{32\eps\sqrt{\gamma^2-1}}
	+\cO(\eps^0)\,,\\
	{\cal I}^{(2;\pm)}_{1,0,1,1,2,0,0}&=
	(2\pi)^{-1+2\eps}e^{-2\eps\gamma_{\rm E}}\frac{i(1+4\eps-8\eps^2)}{16\sqrt{\gamma^2-1}}
	+\cO(\eps^3)\,,\\
	{\cal I}^{(2;\pm)}_{1,0,1,1,1,1,1}&=
	4^{-1-3\eps}\pi^{-1+2\eps}
	e^{-2\eps\gamma_{\rm E}}\frac{i(-1+6\eps)}{8\sqrt{\gamma^2-1}}
	+\cO(\eps^2)\,,
\end{align}
\end{subequations}
which are associated with terms proportional to the velocities $v_i^\mu$.

To derive these expressions for the master integrals
we set up systems of differential equations (DEs).
Using publicly available tools such as
\texttt{Fuchsia}~\cite{Gituliar:2017vzm} and \texttt{epsilon}~\cite{Prausa:2017ltv}
one may find linear transformations to canonical bases $\vec{F}(x,\eps)$
of master integrals that obey
\begin{align}\label{eq:diffEqs}
\frac{\d\vec{F}}{\d x}=\eps\,\mathbb{M}(x)\vec{F}\,,
\end{align}
where $\mathbb{M}(x)$ is a matrix depending only on $x=\gamma-\sqrt{\gamma^2-1}$.
The use of $x$ rather than $\gamma$ in the DEs was proposed in \Rcite{Parra-Martinez:2020dzs}:
$\mathbb{M}(x)$ contains only poles in $\{x,1+x,1-x\}$,
i.e.~the symbol alphabet.\footnote{
	We used \texttt{PolyLogTools}~\cite{Duhr:2019tlz} for manipulation of polylogarithms.
}
The essential property of \Eqn{eq:diffEqs} is factorization of the $\eps$-dependence,
which enables a straightforwards solution to these DEs
for $\vec{F}$ as Laurent series expansions in $\eps=2-\frac{D}2$.
Integration constants are fixed in the potential region
by comparison with the static limit $v\to0$, i.e.~$\gamma\to1$.
To leading order in $v$ the ${\cal I}^{(1;\pm)}$ and ${\cal I}^{(2;\mp)}$
scalar integral families reduce to the same expression:
\begin{align}
	&\{{\cal I}^{(1;\pm)}_{n_1,n_2,\ldots,n_7},{\cal I}^{(2;\mp)}_{n_1,n_2,\ldots,n_7}\}
	\xrightarrow{v\to0}\\
	&\quad\int_{\Bell_1,\Bell_2}\frac{1}
	{(\Bell_1\cdot\mathbf{v}\!+\!i0^+)^{n_1}(\pm\Bell_2\cdot\mathbf{v}\!+\!i0^+)^{n_2}
	(-\Bell_1^2)^{n_3}(-\Bell_2^2)^{n_4}(-(\Bell_1\!+\!\Bell_2\!-\!\mathbf{q})^2)^{n_5}
	(-(\Bell_1\!-\!\mathbf{q})^2)^{n_6}(-(\Bell_2\!-\!\mathbf{q})^2)^{n_7}}
	+\cO(v^{1-n_1-n_2}).\nn
\end{align}
We therefore fix boundary conditions on the ${\cal I}^{(2;\pm)}$
integrals by equating them with the ${\cal I}^{(1;\mp)}$ integrals:
expressions for the masters in this case having already been given~\eqref{eq:easyMasters}.

\sec{4.~Fourier transform}
Finally we perform the Fourier transform
$q$-integrals from \Eqn{eq:integralTemplate}.
These generically take the form
\begin{align}
\begin{aligned}
  I^{(D)}_\nu[q^{\mu_1}q^{\mu_2}\cdots q^{\mu_n}]
  :=\int_qe^{iq\cdot b}\,\dd(q\cdot v_1)\dd(q\cdot v_2)|q|^\nu
  q^{\mu_1}q^{\mu_2}\cdots q^{\mu_n}\,.
\end{aligned}
\end{align}
The scalar integral is well-known --- see e.g.~\Rcite{Mogull:2020sak}:
\begin{align}
  I^{(D)}_\nu=
  \frac{2^\nu}{\pi^{(D-2)/2}\sqrt{\gamma^2-1}}
  \frac{\Gamma(\sfrac{D-2+\nu}2)}{\Gamma(-\sfrac{\nu}2)}
  \left(-b\cdot P_{12}\cdot b\right)^{-\frac{D-2+\nu}2}\!\!,
\end{align}
where $P^{\mu\nu}_{12}:=\eta^{\mu\nu}-w_1^\mu v_1^\nu-w_2^\nu v_2^\nu$
projects to the $(D-2)$-dimensional space orthogonal to $v_i^\mu$.
The generalization to higher-rank integrals follows easily
by taking derivatives with respect to $b^\mu$:
\begin{equation}
  I^{(D)}_\nu[q^{\mu_1}q^{\mu_2}\cdots q^{\mu_n}]=
  (-i)^n\frac{\partial^nI^{(D)}_\nu}
  {\partial b_{\mu_1}\partial b_{\mu_2}\cdots\partial b_{\mu_n}}\,.
\end{equation}
One should avoid imposing $b\cdot v_i=0$
until after these derivatives have been taken ---
hence our use of the projector $P_{12}^{\mu\nu}$.

\section{Results}
\label{sec:results}

In this Appendix we collect our results for the 3PM conservative momentum impulse
$\Delta p_1^{(3)\mu}$ and spin kick $\Delta a_1^{(3)\mu}$,
respectively providing the coefficients $c_i^{(s)\mu}$ and $d_i^{(s)\mu}$
in Eq.~\eqref{eq:schematic3PM}.
At leading (zeroth) order the momentum impulse is well-known (see e.g.~\Rcite{Kalin:2020fhe}):
\begin{subequations}\label{eq:nonSpinImpulse}
\begin{align}
  &c_{0}^{(0)\mu}
  =
  -\hat b^{\mu}\frac{8\left(4 \gamma ^4-12 \gamma ^2-3\right)
  }{\left(\gamma ^2-1\right) }
  \ ,
  \\
  &c_{1}^{(0)\mu}
  =
  \hat b^\mu
  \frac{
    2\left(16\gamma ^6-32\gamma ^4+16\gamma ^2-1\right)
  }{
    \left(\gamma ^2-1\right)^{5/2}}
  +
  (\gamma v_2^\mu-v_1^\mu)
  \frac{3 \pi (5\gamma^2-1)(2\gamma^2-1)
  }{2 \left(\gamma ^2-1\right)^2 }
  \ ,
  \\
  &c_{2}^{(0)\mu}
  =
  \hat b^\mu
  \frac{
    4 \gamma  \left(20 \gamma ^6-90 \gamma ^4+120 \gamma ^2-53\right)}{3 \left(\gamma ^2-1\right)^{5/2}}
  +
  (1+\gamma)(v_2^\mu-v_1^\mu)
  \frac{3 \pi (5\gamma^2-1)(2\gamma^2-1)
  }{2 \left(\gamma ^2-1\right)^2 }
  \ ,
  \\
  &c_{3}^{(0)\mu}
  =
  - \left. c_{1}^{(0)\mu} \right|_{1\leftrightarrow 2}
  \,,
\end{align}
\end{subequations}
and of course there is no spin kick.
The linear-in-spin coefficients of the impulse are
\begin{subequations}
  \begin{align}
    &c_{0}^{(1)\mu}
    =
    \frac{16 \gamma  \left(\gamma ^2-6\right) \left(2 \gamma ^2+1\right)
      (3\hat b^{\mu} l\cdot a_++l^{\mu} \hat b\cdot a_+)}{\left(\gamma ^2-1\right)^{3/2}}\,,
    \\
    &c_{1}^{(1)\mu}
    =
    \hat b^{\mu} \left(-\frac{32 \gamma  \left(3 \gamma ^2-1\right) l\cdot a_1}{(\gamma^2 -1) }-\frac{4 \gamma  \left(40 \gamma ^4-52 \gamma ^2+13\right)
      l\cdot a_2}{(\gamma^2-1)^2 }\right)\\
    &\qquad\quad+l^{\mu} \left(-\frac{4 \gamma  \left(8 \gamma ^4-12 \gamma ^2+3\right) \hat b\cdot a_+}
         {(\gamma^2 -1)^2}
         -\frac{3 \pi  \left(10 \gamma ^4-9 \gamma
           ^2+1\right) a_1\cdot v_2}{2 \left(\gamma ^2-1\right)^{5/2} }+\frac{\pi  \gamma  \left(35 \gamma ^4-40 \gamma ^2+9\right) a_2\cdot v_1}{\left(\gamma
           ^2-1\right)^{5/2} }\right)\nn\\
         &\qquad\quad+(v_1^{\mu}-\gamma v_2^\mu)\left(\frac{3 \pi  \gamma  \left(3 \gamma ^2-1\right) \left(5 \gamma ^2-4\right)
           l\cdot a_1}{\left(\gamma
           ^2-1\right)^{5/2} }+\frac{\pi  \gamma  \left(55 \gamma ^4-62 \gamma ^2+15\right)
           l\cdot a_2}{\left(\gamma ^2-1\right)^{5/2} }\right)\nn\,,
         \\
         &c_{2}^{(1)\mu}
         =
         -\hat b^{\mu}\frac{4 \left(20 \gamma ^6-140
           \gamma ^4+80 \gamma ^2+41\right)  l\cdot a_+}{(\gamma^2-1)^2}
         +l^\mu \Bigg(
         \frac{\pi  \left(60 \gamma ^5+35 \gamma ^4-69 \gamma ^3-30 \gamma ^2+15 \gamma +3\right) a_+\cdot(v_1-v_2)}{2 \left(\gamma ^2-1\right)^{5/2}}
         \\
         &\qquad\quad
         -\frac{8 \left(10 \gamma ^6-78 \gamma ^4+45 \gamma
           ^2+20\right) b\cdot a_+}{3 (\gamma^2-1)^2}
         \Bigg)
         +(v_1^{\mu}l\cdot a_1-v_2^\mu l\cdot a_2)\frac{\pi  \gamma  \left(45 \gamma ^5+55 \gamma ^4-51 \gamma ^3-62 \gamma ^2+12 \gamma +15\right)
         }
         {\left(\gamma ^2-1\right)^{5/2}}
         \nn\\
         &\qquad\quad
         +(v_1^{\mu} l\cdot a_2-v_2^\mu l\cdot a_1) \frac{\pi  \gamma  \left(55 \gamma ^5+45 \gamma ^4-62 \gamma ^3-51 \gamma ^2+15 \gamma +12\right)
         }
         {\left(\gamma ^2-1\right)^{5/2}}\,,\nn
         \\
         &c_{3}^{(1)\mu}
         =
         - \left. c_{1}^{(1)\mu} \right|_{1\leftrightarrow 2}\,,
  \end{align}
\end{subequations}
where $a_+^\mu=a_1^\mu+a_2^\mu$,
$l^\mu:={\eps^\mu}_{\nu\rho\sigma}\hat{b}^\nu v_1^\rho v_2^\sigma/(\gamma v)$ and for brevity
we omit the additional subscripts on $a^\mu_{i0}$.
The coefficients of the spin kick are
\begin{subequations}
  \begin{align}
    &d_{0}^{(1)\mu}
    =
    (v_2^\mu \hat b\cdot a_1-\hat b^{\mu} a_1\cdot v_2)
    \frac{16 \gamma  \left(\gamma ^2-6\right) \left(2 \gamma ^2+1\right)
    }{
      (\gamma^2-1)^2}
    +
    v_1^{\mu} \hat b\cdot a_1 \frac{24 \left(2 \gamma ^4+7 \gamma ^2+1\right) 
    }{
      (\gamma^2-1)^2}
    \ ,\\
    &d_{1}^{(1)\mu}
    =
    \hat b^{\mu} \left(\frac{3 \pi  \hat b\cdot a_1}{2 }
    +
    \frac{8 \gamma  (4 \gamma^2 -1) a_1\cdot v_2}{\left(\gamma ^2-1\right)^{3/2}
      }\right)+v_1^{\mu} \left(\frac{2 \left(16 \gamma ^4-14 \gamma ^2+1\right) \hat b\cdot a_1}{\left(\gamma ^2-1\right)^{5/2} }-\frac{3 \pi  \gamma 
      \left(5 \gamma ^2-2\right) a_1\cdot v_2}{(\gamma^2-1)^2}\right)
    \\&\qquad\quad
    +v_2^{\mu} \left(\frac{3 \pi  \gamma ^2 \left(5 \gamma ^2-3\right) a_1\cdot
      v_2}{(\gamma^2-1)^2 }-\frac{4 \gamma  \left(12 \gamma ^4-14 \gamma ^2+3\right) \hat b\cdot a_1}{\left(\gamma ^2-1\right)^{5/2} }\right)
    \nn\ ,\\
    &d_{2}^{(1)\mu}
    =
    \hat b^{\mu} \left(\frac{\pi  \left(10 \gamma ^4-3 \gamma ^2-3\right) \hat b\cdot a_1}{2 (\gamma^2 -1)}+\frac{4 \left(20 \gamma ^6-132 \gamma ^4+72 \gamma
      ^2+43\right) a_1\cdot v_2}{3 \left(\gamma ^2-1\right)^{5/2} }\right)
    \\&\qquad\quad
    +v_1^{\mu} \left(-\frac{4 \gamma  \left(10 \gamma ^4+20 \gamma ^2-33\right) \hat b\cdot a_1}{\left(\gamma ^2-1\right)^{5/2} }-\frac{\pi 
      \left(30 \gamma ^4+35 \gamma ^3-21 \gamma ^2-15 \gamma +3\right) a_1\cdot v_2}{2 (\gamma^2-1)^2 }\right)
    \nn\\&\qquad\quad
    +v_2^\mu \left(\frac{4 \pi  \gamma
      ^2 \left(5 \gamma ^2-3\right) a_1\cdot v_2}{(\gamma^2-1)^2 }-\frac{8 \left(22 \gamma ^6-84 \gamma ^4+45 \gamma ^2+20\right) \hat b\cdot a_1}{3
      \left(\gamma ^2-1\right)^{5/2} }\right)
    \nn\ ,\\
    &d_{3}^{(1)\mu}
    =
    \hat b^{\mu} a_1\cdot v_2 \frac{4 \gamma  \left(16 \gamma ^4-20 \gamma ^2+5\right) }{\left(\gamma ^2-1\right)^{5/2} }
    -v_2^{\mu} \hat b\cdot a_1
    \frac{4 \gamma  \left(16 \gamma ^4-20 \gamma ^2+5\right) }{\left(\gamma ^2-1\right)^{5/2} }
   \\&\qquad\quad
   +v_1^{\mu} \left(\frac{2 \left(2 \gamma
   ^2+1\right) \left(8 \gamma ^4-8 \gamma ^2+1\right) \hat b\cdot a_1}{\left(\gamma ^2-1\right)^{5/2} }-\frac{3 \pi  \left(2 \gamma ^2-1\right) \left(5 \gamma
     ^2-1\right) a_1\cdot v_2}{2 (\gamma^2-1)^2 }\right)
   \nn\,.
  \end{align}
\end{subequations}
Our quadratic-in-spin results are of a similar nature,
but due to their considerable length
we restrict ourselves to only including spin on the first body,
and with $C_{\rm E,1}=0$
(a Kerr black hole scattering off a Schwarzschild black hole).
For our full results --- including spins on both bodies,
and with finite-size $C_{{\rm E},i}$ coefficients allowing
for a generalization to neutron stars ---
we refer the interested reader to the  ancillary file attached
to the \texttt{arXiv} submission of this Letter.
The components of the momentum impulse are
\begin{subequations}
  \begin{align}
    &c_{0}^{(2)\mu}
    =
    \hat b^{\mu} \bigg(
    \frac{64 \gamma ^2
      \left(4 \gamma ^6-36 \gamma ^4+\gamma ^2+6\right) (\hat b\cdot a_1)^2}{ (\gamma^2 -1)^3}+
    \frac{64 \left(3 \gamma ^8-35 \gamma ^6+9 \gamma
      ^4+42 \gamma ^2+6\right) \left(a_1\cdot v_2\right){}^2}{ (\gamma^2 -1)^4}
    \\&\qquad\quad
    +\frac{192 \left(\gamma ^6-8 \gamma ^4-7 \gamma ^2-1\right)
      a_1^2}{ (\gamma^2-1)^2}\bigg)
    -l^{\mu}\frac{32 \gamma ^2 \left(4 \gamma ^6-36 \gamma ^4+\gamma ^2+6\right) \hat b\cdot a_1 l\cdot a_1 }{ (\gamma^2 -1)^3}\,,
    \nn\\\qquad\quad
    &c_{1}^{(2)\mu}
    =
    \hat b^{\mu} \bigg(
    -\frac{4 \left(64 \gamma ^6-108 \gamma ^4+45 \gamma ^2-2\right) (\hat b\cdot a_1)^2}{\left(\gamma ^2-1\right)^{5/2} }
    -\frac{3 \pi  \gamma
      \left(5 \gamma ^2-2\right) \hat b\cdot a_1 a_1\cdot v_2}{(\gamma^2-1)^2 }
    \\&\qquad\quad
    -\frac{8 a_1^2 \left(24 \gamma ^4-16 \gamma ^2+1\right)}{\left(\gamma
      ^2-1\right)^{3/2} }
    -\frac{4 \left(2 \gamma ^2-1\right) \left(24 \gamma ^4-24 \gamma ^2+1\right) \left(a_1\cdot v_2\right){}^2}{\left(\gamma ^2-1\right)^{7/2}
      }\bigg)
    \nn\\&\qquad
    +l^{\mu} \bigg(
    \frac{4 \left(32 \gamma ^6-60 \gamma ^4+29 \gamma ^2-2\right) \hat b\cdot a_1 l\cdot a_1}{\left(\gamma ^2-1\right)^{5/2}
      }
    +\frac{3 \pi  \gamma  \left(40 \gamma ^4-53 \gamma ^2+15\right) l\cdot a_1 a_1\cdot v_2}{2 (\gamma^2 -1)^3 }
    \bigg)
    \nn\\&\qquad
    +(v_1^{\mu}-\gamma v_2^\mu)
    \bigg(
    \frac{3 \pi  \left(150 \gamma ^6-235 \gamma ^4+101 \gamma ^2-10\right) (\hat b\cdot a_1)^2}{4 (\gamma^2 -1)^3 }
    +\frac{4 \gamma 
      \hat b\cdot a_1 a_1\cdot v_2}{\left(\gamma ^2-1\right)^{7/2} }
    \nn\\&\qquad\quad
    +\frac{3 \pi  a_1^2 \left(60 \gamma ^6-94 \gamma ^4+39 \gamma ^2-3\right)}{2 (\gamma^2 -1)^3 }
    +\frac{3 \pi  \left(120 \gamma ^6-182 \gamma ^4+73 \gamma ^2-5\right) \left(a_1\cdot v_2\right){}^2}{4 (\gamma^2 -1)^4
      }\bigg)\,,
    \nn\\
    &c_{2}^{(2)\mu}
    =
    \hat b^{\mu} \bigg(
    -\frac{8 \gamma  \left(400 \gamma ^6-4528 \gamma ^4+241 \gamma ^2+872\right) (\hat b\cdot a_1)^2}{15  \left(\gamma
      ^2-1\right)^{5/2}}
    -\frac{16 \gamma 
      \left(52 \gamma ^4-514 \gamma ^2-373\right) a_1^2}{5  \left(\gamma ^2-1\right)^{3/2}}
    \\&\qquad\quad
    -\frac{8
      \gamma  \left(216 \gamma ^6-4340 \gamma ^4+2642 \gamma ^2+4497\right) \left(a_1\cdot v_2\right){}^2}{15  \left(\gamma ^2-1\right)^{7/2}}
    -\frac{\pi  \left(60 \gamma ^4+35 \gamma ^3-39 \gamma ^2-15 \gamma +3\right) a_1\cdot v_2 \hat b\cdot a_1}{2  (\gamma^2-1)^2}
    \bigg)
    \nn\\&\qquad
    +l^{\mu}\bigg(\frac{8 \gamma  \left(200 \gamma ^6-2444 \gamma ^4+353
      \gamma ^2+376\right) \hat b\cdot a_1 l\cdot a_1}{15  \left(\gamma ^2-1\right)^{5/2}}
    \nn\\&\qquad\quad
    +\frac{\pi  \left(960 \gamma ^5-295 \gamma ^4-1217 \gamma ^3+327 \gamma ^2+285
      \gamma -36\right) a_1\cdot v_2 l\cdot a_1}{8  (\gamma -1)^3 (\gamma +1)^2}\bigg)
    \nn\\&\qquad
    + v_1^{\mu}\bigg(\frac{\pi  \left(1800 \gamma ^6+710 \gamma ^5-3530 \gamma ^4-489 \gamma ^3+1701 \gamma ^2+63 \gamma -183\right)
      (\hat b\cdot a_1)^2}{16  (\gamma -1)^3 (\gamma +1)^2}+\frac{4 \left(3 \gamma ^2-1\right) a_1\cdot v_2 \hat b\cdot a_1}{ \left(\gamma
      ^2-1\right)^{7/2}}
    \nn\\&\qquad\quad
    +\frac{\pi  \left(1440 \gamma ^6+790 \gamma ^5-2974 \gamma ^4-537 \gamma ^3+1413 \gamma ^2+63 \gamma -123\right) \left(a_1\cdot v_2\right){}^2}{16
       (\gamma -1)^4 (\gamma +1)^3}
    \nn\\&\qquad\quad
    +\frac{\pi  \left(720 \gamma ^6+290 \gamma ^5-1418 \gamma ^4-195 \gamma ^3+663 \gamma ^2+21 \gamma -57\right) a_1^2}{8
      (\gamma -1)^3 (\gamma +1)^2}\bigg)
    \nn\\&\qquad
    +v_2^{\mu}\bigg(-\frac{8 \hat b\cdot a_1 a_1\cdot v_2 \gamma ^3}{ \left(\gamma ^2-1\right)^{7/2}}-\frac{\pi 
      \left(2510 \gamma ^6-710 \gamma ^5-3309 \gamma ^4+489 \gamma ^3+1275 \gamma ^2-63 \gamma -120\right) (\hat b\cdot a_1)^2}{16  (\gamma -1)^3 (\gamma
      +1)^2}
    \nn\\&\qquad\quad
    -\frac{\pi  \left(2230 \gamma ^6-790 \gamma ^5-2721 \gamma ^4+537 \gamma ^3+939 \gamma ^2-63 \gamma -60\right) \left(a_1\cdot v_2\right){}^2}{16  (\gamma
      -1)^4 (\gamma +1)^3}
    \nn\\&\qquad\quad
    -\frac{\pi  \left(1010 \gamma ^6-290 \gamma ^5-1323 \gamma ^4+195 \gamma ^3+489 \gamma ^2-21 \gamma -36\right) a_1^2}{8  (\gamma -1)^3
      (\gamma +1)^2}\bigg)\,,
    \nn\\
    &c_{3}^{(2)\mu}
    =
    \hat b^{\mu} \bigg(-\frac{4 \left(144 \gamma ^6-248 \gamma ^4+110 \gamma ^2-7\right) (\hat b\cdot a_1)^2}{\left(\gamma ^2-1\right)^{5/2} }-\frac{\pi 
      \left(70 \gamma ^4-45 \gamma ^2+3\right) \hat b\cdot a_1 a_1\cdot v_2}{2 (\gamma^2-1)^2 }
    \\&\qquad\quad
    -\frac{8 a_1^2 \left(56 \gamma ^4-40 \gamma
      ^2+3\right)}{\left(\gamma ^2-1\right)^{3/2} }-\frac{4 \left(144 \gamma ^6-240 \gamma ^4+104 \gamma ^2-7\right) \left(a_1\cdot v_2\right){}^2}{\left(\gamma
      ^2-1\right)^{7/2} }\bigg)
    \nn\\&\qquad
    +l^{\mu} \left(\frac{4 \left(48 \gamma ^6-88 \gamma ^4+42 \gamma ^2-3\right) \hat b\cdot a_1 l\cdot a_1}{\left(\gamma ^2-1\right)^{5/2}
      }+\frac{\pi  \left(1330 \gamma ^6-2125 \gamma ^4+876 \gamma ^2-57\right) l\cdot a_1 a_1\cdot v_2}{8 (\gamma^2 -1)^3 }\right)
    \nn\\&\qquad
    +(\gamma v_1^{\mu}-v_2^\mu) \bigg(\frac{\pi\left(2510 \gamma ^6-4019 \gamma ^4+1764 \gamma ^2-183\right) (\hat b\cdot a_1)^2}{16 (\gamma^2 -1)^3
      }+\frac{4\left(2 \gamma ^2-1\right) \hat b\cdot a_1 a_1\cdot v_2}{\left(\gamma ^2-1\right)^{7/2} }
    \nn\\&\qquad\quad
    +\frac{\pi  a_1^2\left(1010 \gamma
      ^6-1613 \gamma ^4+684 \gamma ^2-57\right)}{8 (\gamma^2 -1)^3 }+\frac{\pi\left(2230 \gamma ^6-3511 \gamma ^4+1476 \gamma ^2-123\right)
      \left(a_1\cdot v_2\right){}^2}{16 (\gamma^2 -1)^4 }\bigg)\,.
    \nn
  \end{align}
\end{subequations}
Finally, the coefficients of the quadratic-in-spin part of the 3PM spin kick are
\begin{subequations}
  \begin{align}
    &d_{0}^{(2)\mu}
    =
    \hat b^{\mu}\frac{32 \left(3 \gamma ^8-35 \gamma ^6+9 \gamma ^4+42 \gamma ^2+6\right)  l\cdot a_1 a_1\cdot v_2}{\left(\gamma ^2-1\right)^{7/2}
      }
    +l^{\mu}\frac{32 \left(\gamma ^8-\gamma ^6-8 \gamma ^4-36 \gamma ^2-6\right)  \hat b\cdot a_1 a_1\cdot v_2}{\left(\gamma ^2-1\right)^{7/2} }
    \\&\qquad
    -v_2^{\mu}\frac{32\left(4 \gamma ^6-36 \gamma ^4+\gamma ^2+6\right) \gamma ^2  \hat b\cdot a_1 l\cdot a_1}{\left(\gamma ^2-1\right)^{7/2} }
    - v_1^{\mu}\frac{32 \left(6 \gamma^6+35 \gamma ^4-4 \gamma ^2-12\right) \gamma  \hat b\cdot a_1 l\cdot a_1}{\left(\gamma ^2-1\right)^{7/2} }\,,
    \nn\\&
    d_{1}^{(2)\mu}
    =
    \hat b^{\mu} \left(
    -\frac{3 \pi  \gamma  \left(2 \gamma ^2-1\right) \hat b\cdot a_1 l\cdot a_1}{\left(\gamma ^2-1\right)^{3/2} }
    -\frac{2 \left(48 \gamma ^6-68 \gamma
      ^4+22 \gamma ^2+1\right) l\cdot a_1 a_1\cdot v_2}{(\gamma^2 -1)^3 }\right)
    \\&\qquad
    +l^{\mu} \left(
    -\frac{3 \pi  \gamma  (\hat b\cdot a_1)^2}{2
      \sqrt{\gamma ^2-1} }
    -\frac{2 \left(16 \gamma ^6-36 \gamma ^4+18 \gamma ^2-1\right) \hat b\cdot a_1 a_1\cdot v_2}{(\gamma^2 -1)^3 }
    -\frac{3
      \pi  \gamma  \left(5 \gamma ^2-2\right) \left(a_1\cdot v_2\right){}^2}{\left(\gamma ^2-1\right)^{5/2} }\right)
    \nn\\&\qquad
    +v_1^{\mu} \left(
    \frac{3 \pi  \left(30
      \gamma ^4-29 \gamma ^2+3\right) l\cdot a_1 a_1\cdot v_2}{2 \left(\gamma ^2-1\right)^{5/2} }
    -\frac{8 \gamma  \left(16 \gamma ^2-7\right) \hat b\cdot a_1 l\cdot
      a_1}{(\gamma^2 -1)^2 }\right)
    \nn\\&\qquad
    + v_2^{\mu} \left(
    \frac{8 \left(24 \gamma ^4-16 \gamma ^2+1\right) \hat b\cdot a_1 l\cdot a_1}{(\gamma^2 -1)^2 }
    -\frac{9 \pi  \gamma  \left(10 \gamma ^2-3\right) l\cdot a_1 a_1\cdot v_2}{2 \left(\gamma ^2-1\right)^{3/2} }\right)\,,
    \nn\\&
    d_{2}^{(2)\mu}
    =
    \hat b^{\mu} \left(
    -\frac{\pi 
      \left(65 \gamma ^3-3 \gamma -6\right) \hat b\cdot a_1 l\cdot a_1}{4 \sqrt{\gamma ^2-1} }
    -\frac{4 \gamma  \left(216 \gamma ^6-4220 \gamma ^4+2492 \gamma
      ^2+4557\right) l\cdot a_1 a_1\cdot v_2}{15 (\gamma^2-1)^3 }\right)
    \\&\qquad
    +
    l^{\mu} \Bigg(
    -\frac{\pi  \left(10 \gamma ^5-3 \gamma ^3+3 \gamma ^2-3
      \gamma -3\right) (\hat b\cdot a_1)^2}{2 \left(\gamma ^2-1\right)^{3/2} }
    -\frac{4 \gamma  \left(184 \gamma ^6-188 \gamma ^4-2416 \gamma ^2-3625\right)
      \hat b\cdot a_1 a_1\cdot v_2}{15 (\gamma^2-1)^3 }
    \nn\\&\qquad\quad
    -\frac{\pi  \left(60 \gamma ^4+35 \gamma ^3-39 \gamma ^2-15 \gamma +3\right) \left(a_1\cdot
      v_2\right){}^2}{2 \left(\gamma ^2-1\right)^{5/2} }\Bigg)
    \nn\\&\qquad
    +
    v_1^{\mu} l\cdot a_1\left(
    \frac{16 \left(222 \gamma ^6+1211 \gamma ^4-483 \gamma ^2-200\right) \hat b\cdot a_1
      }{15 (\gamma^2 -1)^3 }
    +\frac{3 \pi  \left(120 \gamma ^5+175 \gamma ^4-176 \gamma ^3-170 \gamma ^2+48 \gamma +19\right) 
      a_1\cdot v_2}{8 \left(\gamma ^2-1\right)^{5/2} }\right)
    \nn\\&\qquad
    +
    v_2^{\mu} l\cdot a_1 \left(
    \frac{16 \gamma  \left(220 \gamma ^6-1342 \gamma ^4+169 \gamma ^2+203\right)
      \hat b\cdot a_1 }{15 (\gamma^2 -1)^3 }
    -\frac{\pi  \left(305 \gamma ^5-120 \gamma ^4-392 \gamma ^3+78 \gamma ^2+87 \gamma -6\right) 
      a_1\cdot v_2}{4 \left(\gamma ^2-1\right)^{5/2} }\right)\,,
    \nn\\&
    d_{3}^{(2)\mu}
    =
    \hat b^{\mu} \bigg(
    \frac{\pi  \left(10 \gamma ^4-3 \gamma ^2-3\right) \hat b\cdot a_1 l\cdot a_1}{2 \left(\gamma ^2-1\right)^{3/2} }
    -\frac{2 \left(160 \gamma ^6-264
      \gamma ^4+116 \gamma ^2-9\right) l\cdot a_1 a_1\cdot v_2}{(\gamma^2 -1)^3 }\bigg)
    \\&\qquad
    +
    l^{\mu} \left(
    -\frac{\pi  \left(10 \gamma ^4-3 \gamma
      ^2-3\right) (\hat b\cdot a_1)^2}{2 \left(\gamma ^2-1\right)^{3/2} }
    +\frac{2 \left(8 \gamma ^4-4 \gamma ^2-1\right) \hat b\cdot a_1 a_1\cdot v_2}{(\gamma^2
      -1)^3 }
    -\frac{\pi  \left(70 \gamma ^4-45 \gamma ^2+3\right) \left(a_1\cdot v_2\right){}^2}{2 \left(\gamma ^2-1\right)^{5/2}
      }\right)
    \nn\\&\qquad
    +
    v_1^{\mu} \bigg(
    \frac{\pi  \gamma  \left(110 \gamma ^4-159 \gamma ^2+45\right) l\cdot a_1 a_1\cdot v_2}{2 \left(\gamma ^2-1\right)^{5/2}
      }
    -\frac{16 \gamma  \left(2 \gamma ^2-1\right) \left(4 \gamma ^2+3\right) \hat b\cdot a_1 l\cdot a_1}{(\gamma^2 -1)^2 }\bigg)
    \nn\\&\qquad
    +v_2^{\mu}
    \left(
    \frac{16 \left(20 \gamma ^4-14 \gamma ^2+1\right) \hat b\cdot a_1 l\cdot a_1}{(\gamma^2 -1)^2 }
    +\frac{\pi  \left(70 \gamma ^4-45 \gamma
      ^2+3\right) l\cdot a_1 a_1\cdot v_2}{2 \left(\gamma ^2-1\right)^{5/2} }\right)\,.
    \nn
  \end{align}
\end{subequations}

\end{widetext}

\bibliographystyle{JHEP}
\bibliography{../bib/wqft_spin}

\end{document}